\documentclass[traditabstract,longauth]{aa}

\usepackage{graphicx}
\usepackage{txfonts}
\usepackage{natbib}
\usepackage{array}

\usepackage{enumitem}

\begin{document}

   \title{The LOFAR Multifrequency Snapshot Sky Survey (MSSS)}
   \titlerunning{LOFAR MSSS}

   \subtitle{I. Survey description and first results}

   \author{G.~H.~Heald\inst{1,2}
   		\and R.~F.~Pizzo\inst{1}
		\and E.~Orr\'u\inst{1}
		\and R.~P.~Breton\inst{3}
		\and D.~Carbone\inst{4}
		\and C.~Ferrari\inst{5}
		\and M.~J.~Hardcastle\inst{6}
		\and W.~Jurusik\inst{7}
		\and G.~Macario\inst{5}
		\and D.~Mulcahy\inst{8,3}
% ...
\email{heald@astron.nl}}

   \institute{ASTRON, the Netherlands Institute for Radio Astronomy, Postbus 2, 7990 AA, Dwingeloo, The Netherlands \and 
   			  Kapteyn Astronomical Institute, University of Groningen, PO Box 800, 9700 AV, Groningen, The Netherlands \and 
			  School of Physics and Astronomy, University of Southampton, SO17 1BJ, UK \and 
			  Anton Pannekoek Institute, University of Amsterdam, Postbus 94249, 1090 GE Amsterdam, The Netherlands \and 
			  Laboratoire Lagrange, UMR 7293, Universit\'e de Nice Sophia-Antipolis, CNRS, Observatoire de la C\^ote d'Azur, 06300 Nice, France \and 
			  School of Physics, Astronomy and Mathematics, University of Hertfordshire, College Lane, Hatfield AL10 9AB, UK \and 
			  Astronomical Observatory, Jagiellonian University, ul. Orla 171, 30-244, Krak\'ow, Poland \and 
			  Max-Planck-Institut f\"ur Radioastronomie, Auf dem H\"ugel 69, 53121, Bonn, Germany
% ...
}

   \date{Received date / Accepted date}

  \abstract{
  We present the Multifrequency Snapshot Sky Survey (MSSS), the first northern-sky LOFAR imaging survey. In this introductory paper, we first describe in detail the motivation and design of the survey. Compared to previous radio surveys, MSSS is exceptional due to its intrinsic multifrequency nature providing information about the spectral properties of the detected sources over more than two octaves (from 30 to 160~MHz). The broadband frequency coverage, together with the fast survey speed generated by LOFAR's multibeaming capabilities, make MSSS the first survey of the sort anticipated to be carried out with the forthcoming Square Kilometre Array (SKA). Two of the sixteen frequency bands included in the survey were chosen to exactly overlap the frequency coverage of large-area Very Large Array (VLA) and Giant Metrewave Radio Telescope (GMRT) surveys at 74\,MHz and 151\,MHz respectively. The survey performance is illustrated within the ``MSSS Verification Field'' (MVF), a region of 100 square degrees centered at $(\alpha,\delta)_\mathrm{J2000}=(15^\mathrm{h},69^\circ)$. The MSSS results from the MVF are compared with previous radio survey catalogs. We assess the flux and astrometric uncertainties in the catalog, as well as the completeness and reliability considering our source finding strategy. We determine the 90\% completeness levels within the MVF to be 100 mJy at 135~MHz with $108\arcsec$ resolution, and 550 mJy at 50~MHz with $166\arcsec$ resolution. Images and catalogs for the full survey, expected to contain 150\,000--200\,000 sources, will be released to a public web server. We outline the plans for the ongoing production of the final survey products, and the ultimate public release of images and source catalogs.
  }

  \keywords{Surveys --- Radio continuum: general}

\maketitle

\section{Background}

All-sky continuum surveys are a key application of radio telescopes. They provide a view of galaxies across the Universe free from the biasing effect of extinction, enable searches for rare sources, and provide a pathway for the discovery of new phenomena. Several large-area surveys have been performed with existing radio telescopes at a number of frequencies. Among these, many of the earliest were performed at low frequencies ($\nu\lesssim350\,\mathrm{MHz}$) \citep[see e.g.][]{jauncey_1975}. The new Low Frequency Array \citep[LOFAR;][]{vanhaarlem_etal_2013} operates at frequencies between 10 and 240~MHz, and was constructed with one primary aim being to perform groundbreaking imaging surveys of the northern sky \citep{rottgering_etal_2011}. Currently, the most extensive low frequency catalogs at or near LOFAR frequencies are the Eighth Cambridge Survey of Radio Sources \citep[8C;][catalog revised by \citealt{hales_etal_1995}]{rees_1990}, the VLA Low-Frequency Sky Survey \citep[VLSS;][and the revised catalog VLSSr; \citealt{lane_etal_2012}]{cohen_etal_2007}, the Seventh Cambridge Survey of Radio Sources \citep[7C;][]{hales_etal_2007}, the Westerbork Northern Sky Survey \citep[WENSS;][]{rengelink_etal_1997}, and most recently the Murchison Widefield Array \citep[MWA;][]{tingay_etal_2013} Commissioning Survey \citep[MWACS;][]{hurley-walker_etal_2015}. Other ongoing low-frequency radio surveys include the TIFR GMRT Sky Survey (TGSS\footnote{\tt http://tgss.ncra.tifr.res.in/150MHz/tgss.html}) that is using the Giant Metrewave Radio Telescope (GMRT) and has already released survey data products to the community, and the Galactic and Extragalactic MWA Survey \citep[GLEAM;][]{wayth_etal_2015}.

A key application of all-sky radio surveys is the comparison of source properties at the wide range of frequencies at which they are detected. This provides crucial information from which the physical properties of these sources can be derived.

To date, no wide-area radio surveys have been performed with large fractional bandwidth (i.e., 2:1 or more). This situation is bound to change in the era of the Square Kilometre Array \citep[SKA;][]{carilli_rawlings_2004}. The SKA and its pathfinder and precursor projects plan wide-area surveys of radio continuum sources, with large fractional bandwidth. This opens the door for the spectral study of sources detected within the survey, using only the survey data itself. LOFAR is a key SKA pathfinder telescope \citep{vanhaarlem_etal_2013} in the 10 and 240~MHz frequency range. The array is centered in the Netherlands with current outlying stations in Germany, France, the United Kingdom, and Sweden. LOFAR is built up from thousands of dipoles clustered in groups called stations. The signals from the dipoles making up each station are digitally combined to steer the beam in one or more directions of interest. Stations are combined in a software correlator located in Groningen, a city in the north of the Netherlands.

One of the key applications of LOFAR is wide-field imaging. In this paper we introduce a new radio survey performed with LOFAR, the Multifrequency Snapshot Sky Survey (MSSS), that has been driven forward as a commissioning project for the telescope. The MSSS project serves as a testbed for operations -- particularly large-scale imaging projects -- and enables straightforward processing of later observations.

The motivations for performing MSSS and the design of the survey are described in detail in \S\,\ref{section:design}. The calibration and imaging strategies are presented \S\,\ref{section:calibration}, and the resulting standard data products are described in \S\,\ref{section:products}. We illustrate the performance of the survey through a detailed analysis of the ``MSSS Verification Field'' (MVF) in \S\,\ref{section:mvf}. Several avenues for scientific exploitation of MSSS are outlined in \S\,\ref{section:science}, and we conclude the paper in \S\,\ref{section:conclusion}.

\section{Context and survey design}\label{section:design}

Imaging applications with the LOFAR telescope will require automated processing. The calibration step in particular needs to be largely unattended, with a major implication that {\it a priori} sky models are required at arbitrary locations on the sky. A number of northern sky radio surveys are available in the literature, but do not cover the proper frequency range at the resolution needed to reliably initiate the calibration routines. Moreover, a coherent commissioning project was required to focus development activities and produce a generic automated processing pipeline, while simultaneously exercising the end-to-end telescope operations. These goals led to the initiation of the ``Multifrequency Snapshot Sky Survey'' (MSSS\footnote{The original name of the survey was the ``Million Source Shallow Survey''. Under current projections (see \S\,\ref{section:mvf}), we expect to catalog well over $10^5$ sources, but probably not $10^6$.}).

In the original MSSS plans (early 2008), it was anticipated that the survey would be performed using 13 core stations (CS), 7 remote stations (RS), and 3 international stations (IS). Ultimately, the array construction proceeded rapidly, and MSSS has been performed with the full complement of stations (except in HBA; see \S\,\ref{subsection:hbasetup}). This leads to a significantly different array than originally envisioned, both in terms of sensitivity and $uv$ coverage. A complete overview of the LOFAR system is provided by \citet{vanhaarlem_etal_2013}. The telescope layout, as well as processing software limitations (now substantially reduced), led to plans for a low-resolution survey initially, with aspirations for a higher resolution survey in future. This paper represents the initial low-resolution effort, and we present prospects for our plans for higher resolution data products in \S\,\ref{subsection:hires}.

The MSSS survey effort needs to provide images and catalogs with sufficient angular resolution to reliably initiate the self-calibration cycle in the imaging pipeline. At the same time, the frequency coverage needs to be sufficient to ensure that spectral variations within the model are accounted for. These requirements were balanced with the need for a relatively rapid survey, taking on the order of weeks of telescope time to perform.

With all of these considerations in mind, we designed a two-component observational strategy. The Low Band Antenna (LBA) component covers the 30--75\,MHz range, and the High Band Antenna (HBA) component covers the 119--158\,MHz range. The exact frequencies were chosen to evenly cover the LBA range, and to avoid major radio frequency interference (RFI) in the HBA range (see Table \ref{table:surveyconfig}). The number of frequency bands (eight $2\,\mathrm{MHz}$ bands in each of the LBA and HBA components) were chosen to allow multiplexing the sky coverage. In early survey test observations, the ``16-bit'' correlator mode allowed three fields to be observed simultaneously, each with $16\,\mathrm{MHz}$ bandwidth. Near the end of 2012 (when most of the LBA test observations were complete, and test HBA observations were beginning), the ``8-bit'' correlator mode became available, doubling the number of fields that can be simultaneously observed (each with $16\,\mathrm{MHz}$ bandwidth) to six. All observations provide data in all four Stokes parameters. The key parameters for the two frequency components of the survey are summarized in Table \ref{table:surveyconfig}, and the setup of these are described in turn below. 

\subsection{Setup of MSSS-LBA}\label{subsection:lbasetup}

The LBA component of MSSS is observed using the LBA\_INNER configuration. In this mode, the digital station beams are formed using signals from the inner 48 dipoles of each 96-dipole station. The resulting compact station, with diameter $D=32.25$~m, provides a large field of view. International stations are included in all MSSS-LBA observations in addition to the Dutch stations, and these come with the full complement of 96 dipoles. The LBA survey pointings were designed using a nominal primary beam half-power beam width (HPBW) at $60\,\mathrm{MHz}$ of 11\fdg55 (from $\mathrm{HPBW}=\alpha_b\lambda/D$, using $\alpha_b=1.3$ and $\lambda=5$~m). We now know that the appropriate value of $\alpha_b=1.10\pm0.02$ \citep{vanhaarlem_etal_2013}, so the station beams are $15\%$ smaller than initially anticipated, and a somewhat larger variation in image noise across the survey can therefore be expected. This will be far less evident at MSSS-LBA frequencies lower than $60\,\mathrm{MHz}$, where the beam sizes are much larger.
Survey fields are laid out on strips of constant declination, and spaced equally on each strip by $\Delta\alpha\leq\mathrm{HPBW}/2$. The declination strips are themselves separated by $\Delta\delta=\mathrm{HPBW}/2$. This results in a total of 660 MSSS-LBA fields.

\begin{table*}
\caption{LOFAR MSSS configuration}
\label{table:surveyconfig}
\centering
\begin{tabular}{lll}
\hline\hline
Parameter & MSSS-LBA & MSSS-HBA \\
\hline
Station configuration & LBA\_INNER & HBA\_DUAL\_INNER \\
Band central frequencies (MHz) & \ & \ \\
~~~Band 0 & 31 & 120 \\
~~~Band 1 & 37 & 125 \\
~~~Band 2 & 43 & 129 \\
~~~Band 3 & 49 & 135 \\
~~~Band 4 & 54 & 143 \\
~~~Band 5 & 60 & 147 \\
~~~Band 6 & 66 & 151 \\
~~~Band 7 & 74 & 157 \\
Calibrator observation type & Simultaneous & Serial \\
Calibrator observation time (min) & 11 & 1 \\
Observing time per snapshot (min) & 11 & 7 \\
Number of snapshots & 9 & 2 \\
Snapshot gap (h) & 1 $(\delta>30^\circ)$ & 4 \\
\  & 0.75 $(\delta<30^\circ)$ & \  \\
Number of fields & 660 & 3616 \\
Total survey time with overheads (h) & 297 & 201 \\
\hline
\end{tabular}
\end{table*}

Initial tests, performed before the station digital beam forming was properly calibrated, indicated that interleaving calibrator observations between target observations would not lead to a sufficiently stable amplitude scale (see \S\,\ref{section:calibration} for details of the calibration strategy). Thus, we adopted an observing mode wherein one of the primary calibrators is always observed, using the identical $16\,\mathrm{MHz}$ bandwidth, in parallel with two simultaneous target field observations (or five target fields, when the 8-bit mode is used).

The amount of observing time used per field is based on the goal of obtaining image noise at the $10\,\mathrm{mJy\,beam}^{-1}$ level, in each of the eight 2~MHz bands. Using early projections of the image sensitivity expected from LOFAR, an estimate of 90 minutes per field was obtained. To improve $uv$ coverage (see below) this was split up into 9 snapshots. The start times of the individual snapshots are spaced by 1~h for northern target fields ($\delta>30\degr$), and by 45~min for target fields closer to the celestial equator ($\delta<30\degr$). The central (fifth) snapshot is observed near transit ($\mathrm{HA}\approx0~\mathrm{h}$). For ease of scheduling, the final survey observing time per field is $9\times11=99\,\mathrm{min}$, yielding initially estimated theoretical thermal noise values between about 6--20~$\mathrm{mJy\,beam}^{-1}$ depending on band.

\begin{figure}
\centering
\includegraphics[width=0.9\hsize]{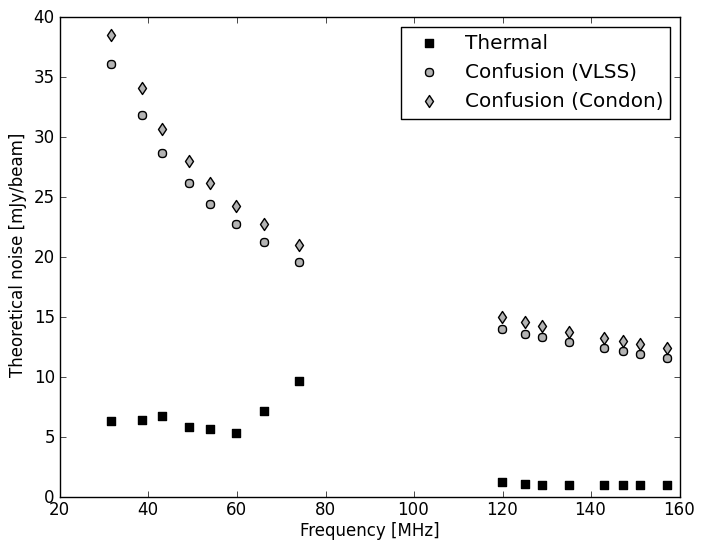}
\caption{Theoretical thermal and confusion noise per band for MSSS observations. The values take into account empirically-determined SEFD values from \citet{vanhaarlem_etal_2013}, and a $uv$ distance cutoff of $3\,\mathrm{k}\lambda$.}
\label{figure:thnoise}
\end{figure}

From observations of calibrator sources, we now have empirical estimates of the LOFAR station System Equivalent Flux Densities (SEFDs); these are provided by \citet{vanhaarlem_etal_2013}. Based on those numbers we can see that an observing time of 99~min yields an expected thermal noise between about 5--10~$\mathrm{mJy\,beam}^{-1}$. It should be noted however that the rudimentary calibration procedures that are implemented in the MSSS pipeline limit our actual sensitivity to an image noise that is typically a factor of a few higher than the thermal noise estimate. Moreover, classical confusion noise would likely limit images produced at the limited angular resolution targeted for default MSSS imaging (out to $uv$ distance of 2--3~$\mathrm{k}\lambda$), see Fig.~\ref{figure:thnoise}. We calculate the expected confusion noise in two ways. The first is based on extrapolation from VLSS B-configuration estimates of the confusion noise \citep[see][]{cohen_2004}:
\begin{equation}
\sigma_{\mathrm{conf,VLSS}}=29\,\left(\frac{\theta}{1\arcsec}\right)^{1.54}\left(\frac{\nu}{74\,\mathrm{MHz}}\right)^{-0.7}\,\mu\mathrm{Jy\,beam^{-1}},
\end{equation}
where $\theta$ is the synthesized beamsize, $\nu$ is the observing frequency, and we have extrapolated the VLSS estimate using a typical spectral index of $-0.7$. The second estimate is from \citet{condon_etal_2012}, who used deep VLA C-configuration observations at S-band ($2-4\,\mathrm{GHz}$) to derive
\begin{equation}
\sigma_{\mathrm{conf,Condon}}=1.2\,\left(\frac{\theta}{8\arcsec}\right)^{10/3}\left(\frac{\nu}{3.02\,\mathrm{GHz}}\right)^{-0.7}\,\mu\mathrm{Jy\,beam^{-1}}.
\end{equation}
We consider the numbers predicted for MSSS fairly reliable since the two estimates provide very similar values (on average only $6.4\%$ different across the full MSSS frequency range) despite being based on data from very different observing frequencies and angular resolutions. Still, the sensitivity obtained in the MSSS data presented in \S\,\ref{section:mvf} suggests that the confusion limit is somewhat lower than predicted, at least in the HBA.

\begin{figure*}
\centering
\includegraphics[width=0.45\hsize]{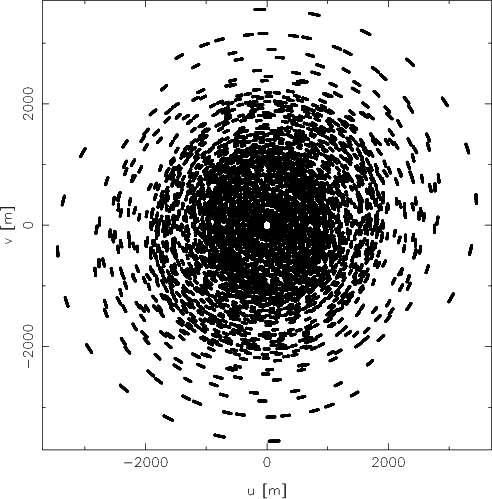}\,\includegraphics[width=0.45\hsize]{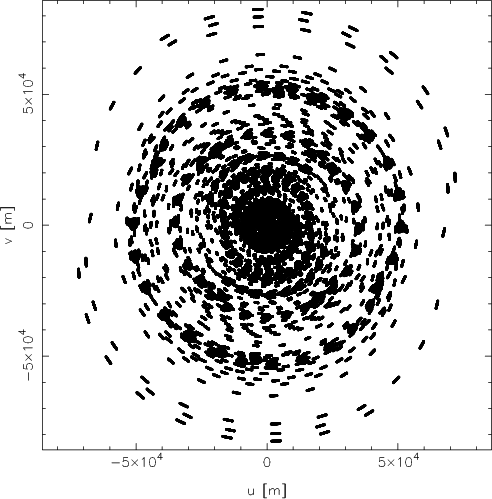}
\caption{The $uv$ coverage for the MSSS-LBA field L227+69, core only (left) and all Dutch stations (right).}
\label{figure:uvcoverage_lba}
\end{figure*}

The $uv$ coverage obtained by splitting the observations of LBA fields into 9 snapshots is illustrated in Fig.~\ref{figure:uvcoverage_lba}. The figure was created using the observations described in \S\,\ref{section:mvf}. 

Including overheads, the total amount of observing time required to complete MSSS-LBA (assuming 8-bit mode) is 297~h. Approximately 130~TB of raw visibility data is collected by observing all 660 pointings and associated calibrator scans.

During the LBA portion of the survey, a single subband (bandwidth 195~kHz) centered at 60~MHz is always placed on the north celestal pole (NCP) as part of a transient and variability monitoring campaign coordinated by the LOFAR Transients Key Project. We return to this feature of the survey setup in \S\,\ref{section:science}.

\subsection{Setup of MSSS-HBA}\label{subsection:hbasetup}

The HBA component of MSSS is observed using the HBA\_DUAL\_INNER station configuration. In this mode, both 24-tile substations are utilized separately for each of the core stations. The outer 24 tiles of the 48-tile remote stations are disabled so that the field of view of the stations are all identical \citep[at a small cost in sensitivity on the longer baselines; see][]{vanhaarlem_etal_2013}. The survey pointings were designed using a nominal beam size at $150\,\mathrm{MHz}$ of HPBW\,=\,4\fdg84 (using $\alpha_b=1.3$, $\lambda=2$~m and $D=30.75$~m). As in the LBA portion of the survey, our understanding of the station beam sizes has now been empirically determined to be smaller \citep[$\alpha_b=1.02\pm0.01$ for HBA core stations;][]{vanhaarlem_etal_2013}.
International stations are not included in the HBA portion of the survey, because at the time of observations the system had restrictions on the number of correlator inputs, requiring the loss of core stations when international stations were included. Since MSSS is primarily a low angular resolution survey, we kept the full complement of core stations in the survey observations. A separate and complementary survey including the international stations has been conducted to search a representative portion of the LOFAR HBA sky for compact ($\lesssim1\arcsec$) calibrators \citep{moldon_etal_2015}.

With the HBA observations, a simultaneous calibrator strategy as implemented in MSSS-LBA is not required, since the stability is much better. Moreover, such a strategy would not be possible, because the analog beamformer in each of the 16-dipole tiles reduces the field of view to approximately $\mathrm{HPBW}_\mathrm{tile}=\lambda/D_\mathrm{tile}=22.9\,\mathrm{deg}$ (where $D_\mathrm{tile}=5\,\mathrm{m}$ is the tile size) at $150\,\mathrm{MHz}$. Thus, most field observations would not be able to include a parallel primary calibrator observation with sufficient sensitivity. Calibrator observations are therefore performed between field observations, using the same bandwidth as the field observations.

Similar sensitivity considerations as for the LBA component of MSSS led to a required observing time per HBA field of approximately 15~min. Snapshots are also used for HBA observations to somewhat improve $uv$ coverage. Two snapshots are used, with the start times of the snapshots separated by 4~h (bracketing transit, such that the hour angles of the snapshots are $\mathrm{HA}\approx\pm2~\mathrm{h}$). For ease of scheduling we adopted $2\times7\,\mathrm{min}$ integrations per field. This leads to an estimated thermal noise of about $1\,\mathrm{mJy\,beam}^{-1}$ per band, using the empirical SEFD values given by \citet{vanhaarlem_etal_2013}. As with the LBA sensitivity, the actual noise level in HBA images is a factor of a few higher than the thermal noise estimate, and confusion is likely the true limiting factor in images with limited angular resolution, see Fig.~\ref{figure:thnoise}.

\begin{figure*}
\centering
\includegraphics[width=0.45\hsize]{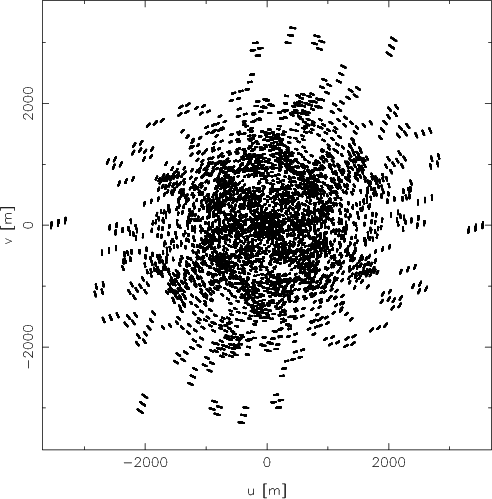}\,\includegraphics[width=0.45\hsize]{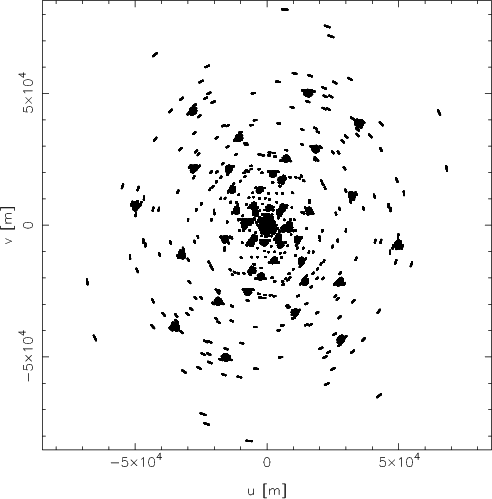}
\caption{The $uv$ coverage for the MSSS-HBA field H229+70, core only (left) and all Dutch stations (right).}
\label{figure:uvcoverage_hba}
\end{figure*}

The $uv$ coverage of the two-snapshot HBA observing strategy is shown in Fig.~\ref{figure:uvcoverage_hba}. The survey grid was designed in the same way as the LBA grid, using HPBW=4\fdg84. The HBA component of MSSS is made up of 3616 fields. Including overheads, the total amount of observing time required to complete MSSS-HBA (assuming 8-bit mode) was 201~h. Approximately 470~TB of raw visibility data is collected by observing all 3616 pointings and associated calibrator scans.

\subsection{Survey fields}\label{subsection:fields}

\begin{figure}
\includegraphics[width=\hsize]{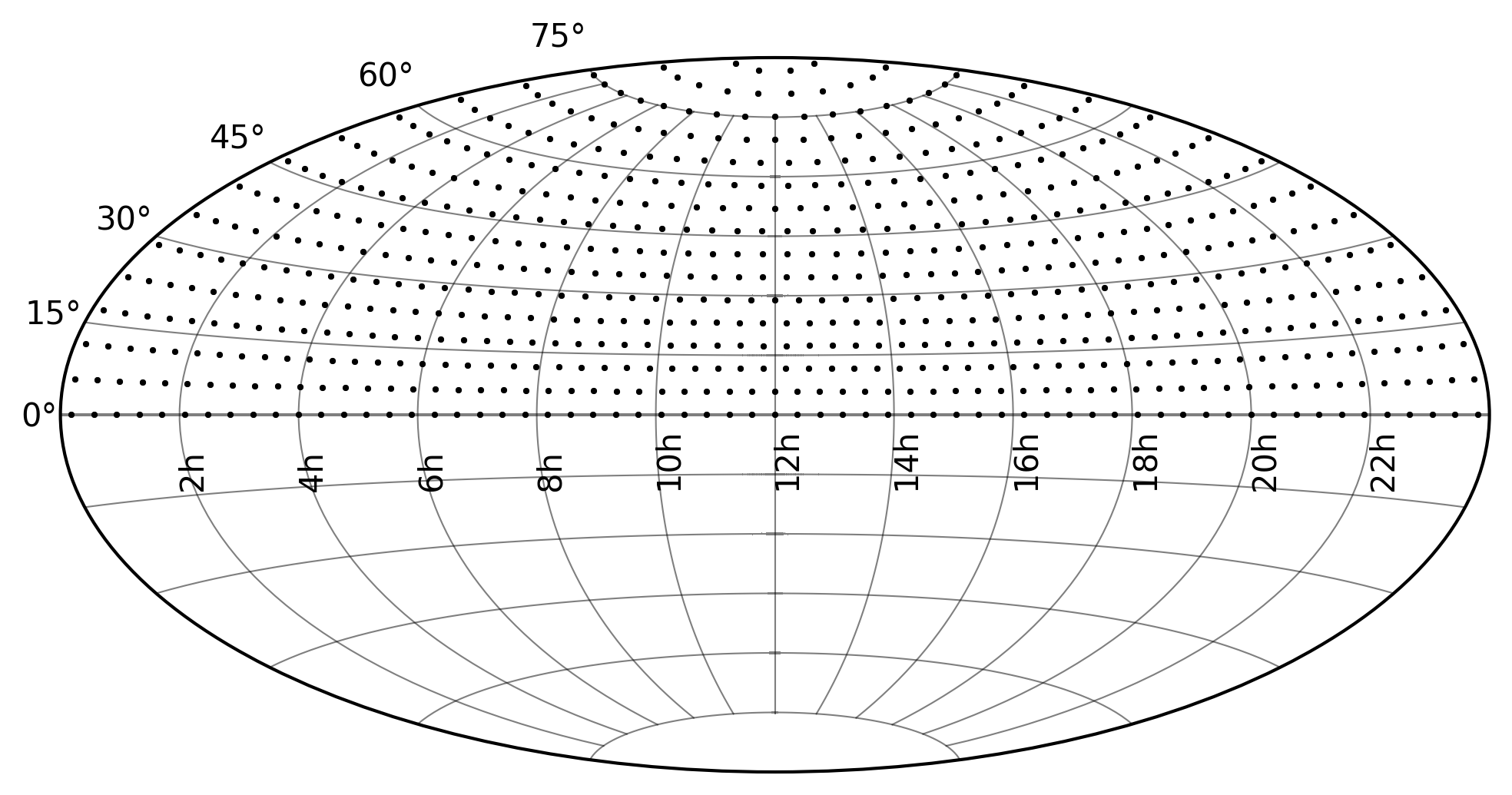}\\
\includegraphics[width=\hsize]{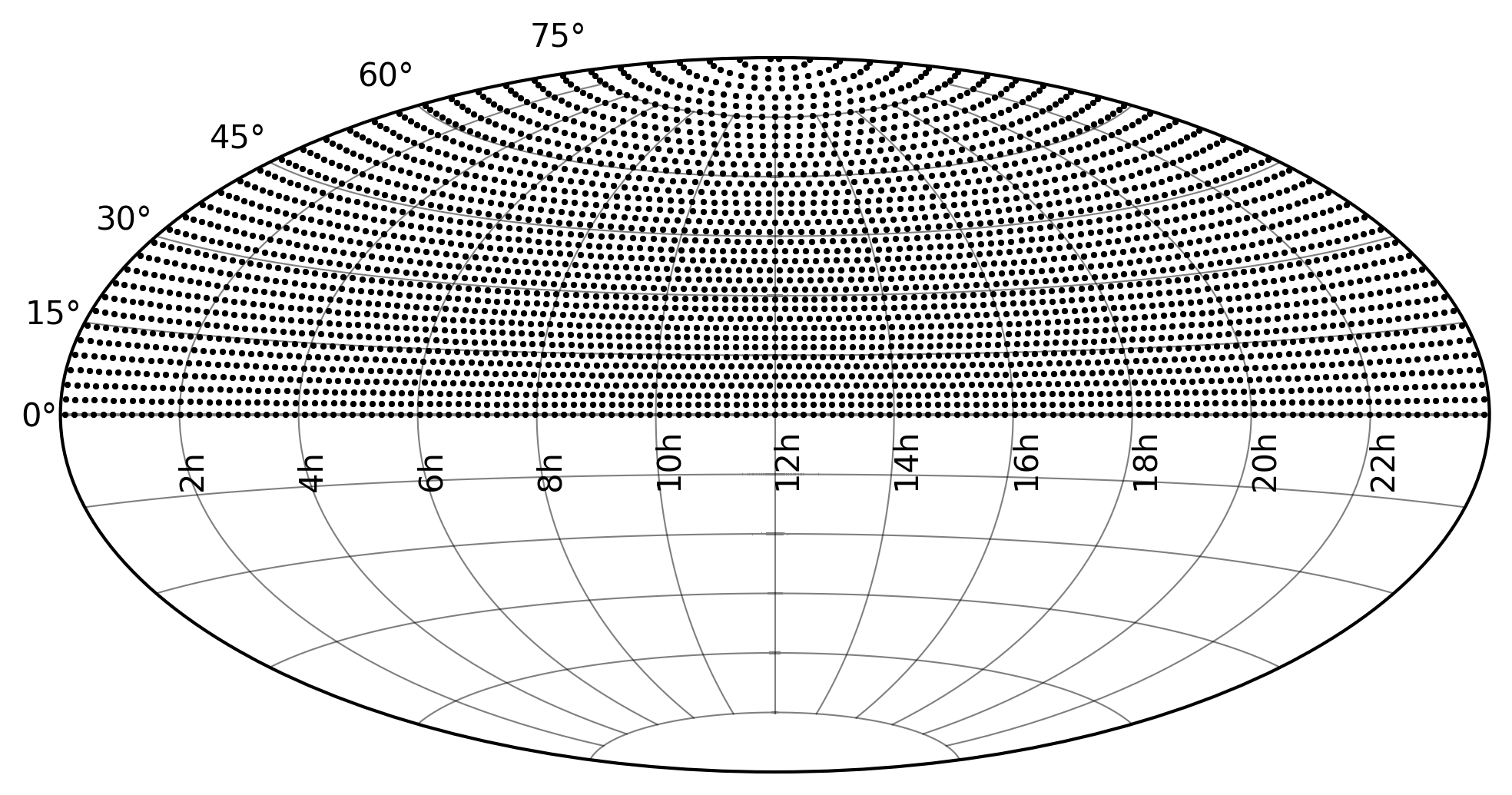}
\caption{MSSS fields for the LBA ({\it top}) and HBA ({\it bottom}) portions of the survey, presented in equatorial coordinates on an Aitoff projection. The LBA survey consists of 660 fields, while the HBA survey consists of 3616 fields.}
\label{figure:grids}
\end{figure}

The layout of the MSSS survey fields was determined in such a way as to provide nearly uniform coverage at the optimized frequencies 60~MHz (MSSS-LBA) and 150~MHz (MSSS-HBA), as described in \S\,\ref{subsection:lbasetup} and \ref{subsection:hbasetup}. The coordinates of the center of the survey fields are shown on an Aitoff projection in Fig.~\ref{figure:grids}.

The MSSS fields planned and observed so far have a lower declination limit of $\delta=0^\circ$. At a later date, the lower declination limit may be extended farther to the south, extending the MSSS sky coverage. 

\subsection{Survey mosaics}\label{subsection:mosaicgrid}

The final presentation of the survey images (and derivation of the resulting catalog) will be based on $10\degr\times10\degr$ mosaics generated from the individual HBA and LBA fields. The mosaic grid is common for both bands in order to facilitate multi-frequency flux comparison of sources detected in the survey. The survey contains a total of 214 mosaic fields, including one NCP mosaic centered at $\delta=90\degr$.

\section{Calibration and imaging strategy}\label{section:calibration}

The MSSS commissioning project has driven the development of the first version of the standard imaging pipeline \citep[SIP;][]{heald_etal_2010}, which can be scheduled by the control system to run automatically on the central processing cluster upon completion of the individual observations. The SIP embodies the calibration strategy described in this section, and is now being upgraded to allow improved image quality at higher angular resolution.

\begin{figure*}
  \centering
  \resizebox{0.75\hsize}{!}{\includegraphics{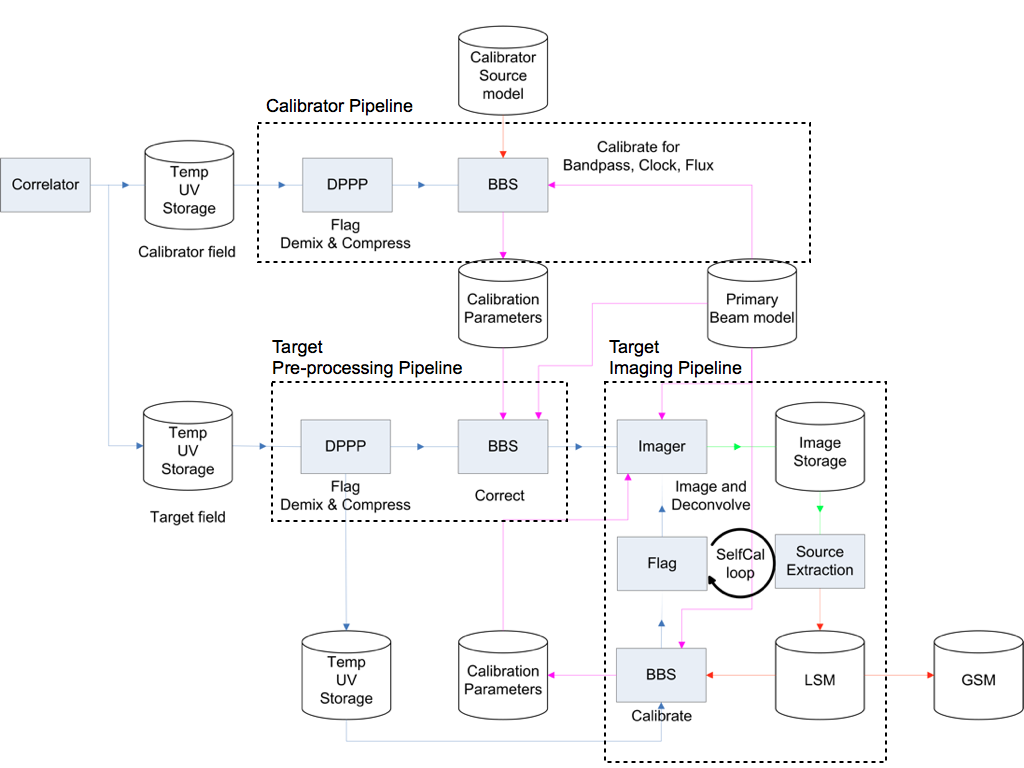}}
  \caption{Sketch of the MSSS calibration and imaging pipeline. See the text for a description.}
  \label{figure:pipeline}
\end{figure*}

\begin{figure}
  \resizebox{\hsize}{!}{\includegraphics{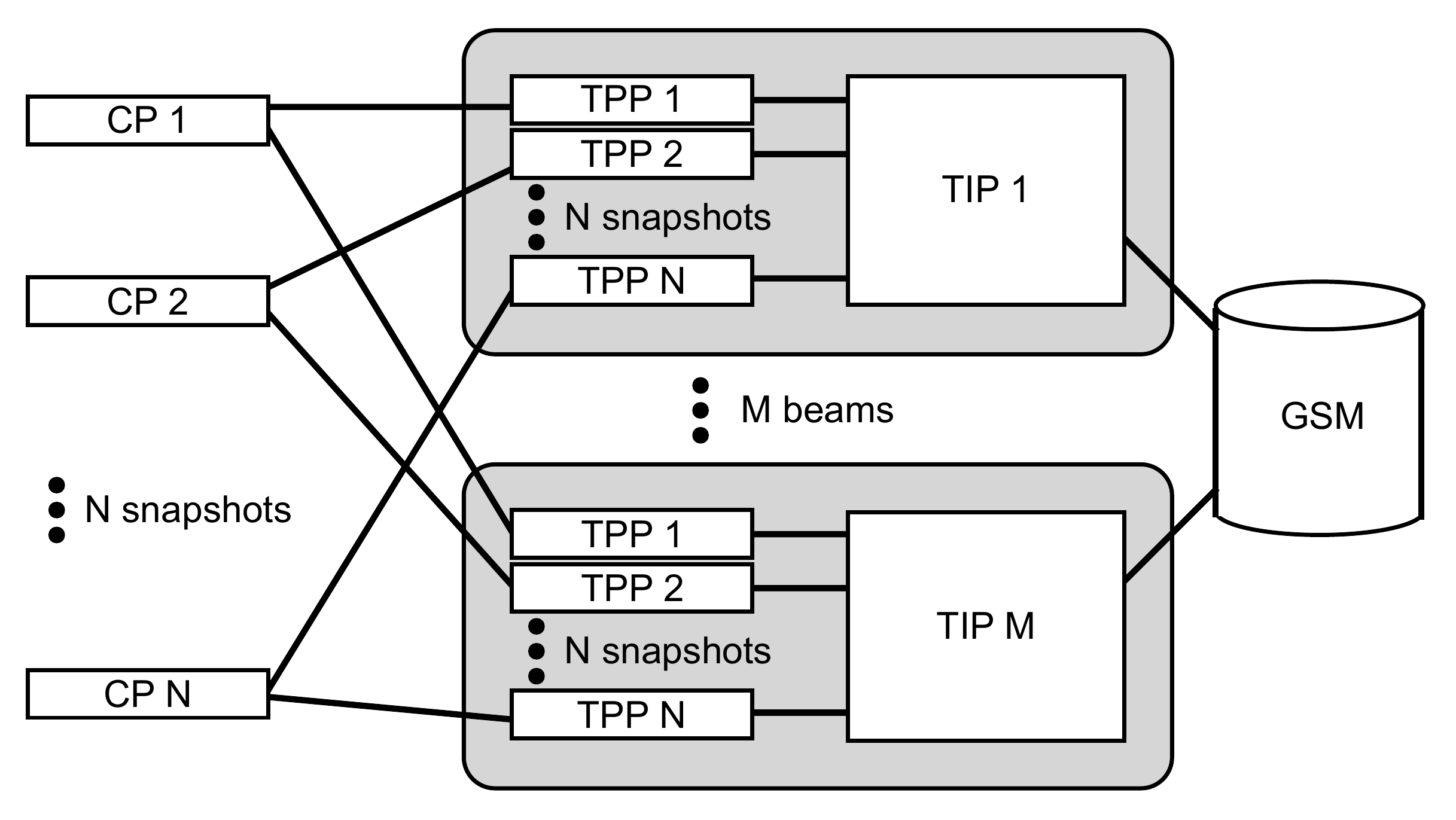}}
  \caption{The scheme for calibrating, processing, and combining individual snapshots of an MSSS field. Each field is observed in $N$ snapshots, and each snapshot observation simultaneously observes $M$ beam directions. In the case of MSSS-LBA, $N=9$, and $M$ is either 2 or 5 depending on whether the observations are done in 16-bit or 8-bit mode. For MSSS-HBA, $N=2$ and $M=3$ or 6. CP stands for Calibrator Pipeline; TPP for Target Pre-processing Pipeline; TIP for Target Imaging Pipeline; and GSM for Global Sky Model. See Fig.~\ref{figure:pipeline} for an overview of how these pieces fit together in more detail, and for the steps that make up each segment of the full pipeline.}
  \label{figure:highlevelcal} 
\end{figure}

We illustrate the processing chain followed by a single observation in Fig.~\ref{figure:pipeline}. Fig.~\ref{figure:highlevelcal} shows a schematic view of the overall processing pipeline, which combines multiple snapshots and calibration observations to create final combined images and feeds source catalogs into the LOFAR Global Sky Model (GSM) database. In these diagrams, cylinders indicate data products, and blocks indicate programs and pipeline segments. The full pipeline is broken up into three main components. The first is the Calibrator Pipeline (CP), which applies some pre-processing steps that are described below to the calibrator scans, and performs primary calibration using a known and well-understood calibrator model. The Target Pre-processing Pipeline (TPP) performs the same pre-processing steps as in CP, and subsequently uses the station gains derived in CP to apply the primary calibration to the individual field snapshots. The resulting calibrated snapshots are stored until all snapshots of a particular field are completed, after which the Target Imaging Pipeline (TIP) begins. This final stage is the heart of the pipeline, and consists of imaging the field and optionally running a self-calibration cycle which is still undergoing further development. We now proceed to detail the various pieces of the pipeline.

\subsection{Pre-processing steps}

Flagging for RFI is performed in a standard and automatic fashion using the AOFlagger \citep{offringa_etal_2010}. This program has been shown to provide excellent RFI excision with minimal false positives. Details of the typical performance of the implemented algorithm on LOFAR data, along with representative RFI statistics, are presented by \citet{offringa_etal_2013}.

Following the flagging step, the demixing technique \citep{vdtol_etal_2007} is applied in order to remove far off-axis, bright sources (primarily the so-called ``A-team'': Cygnus A, Cassiopeia A, Virgo A, and Taurus A) from the visibility data. The automatic pipeline calculates the distance to the A-team sources from the target (and calibrator) fields. Sources within distance ranges determined empirically from commissioning experience are selected for demixing. 

Compression of the data to a manageable volume is done automatically following the demixing step. Typically, in both LBA and HBA the data volume is reduced by a factor of 10 after averaging and calibration have been performed.
The compression factors were selected to minimize bandwidth and time smearing effects, as well as retain sufficient time resolution to allow capturing time variable ionospheric effects. In this way we are able to restrict our estimates of position-dependent survey sensitivity to only consider the combination of station beam pattern and survey pointing grid.

\subsubsection{Bandwidth smearing}

The effect of finite bandwidth is to partially decorrelate the signal and leads to a radial smearing of sources far from the phase tracking center. Assuming a square bandpass and a synthesized beam with a Gaussian profile, the magnitude of the reduction in peak flux density can be approximated as given by \citet{bridle_schwab_1999}:
\begin{equation}
\frac{I}{I_0}=\frac{\sqrt{\pi}}{2\sqrt{\ln{2}}}\frac{\theta\nu_\mathrm{c}}{r\Delta\nu}\mathrm{erf}\left(\sqrt{\ln{2}}\frac{r\Delta\nu}{\theta\nu_\mathrm{c}}\right)
\end{equation}
where $\theta$ is the synthesized beam size (FWHM), $\nu_\mathrm{c}$ is the central frequency of the observation, $r$ is the angular distance from the phase center, and $\Delta\nu$ is the bandwidth. 

\begin{figure}
\includegraphics[width=\hsize]{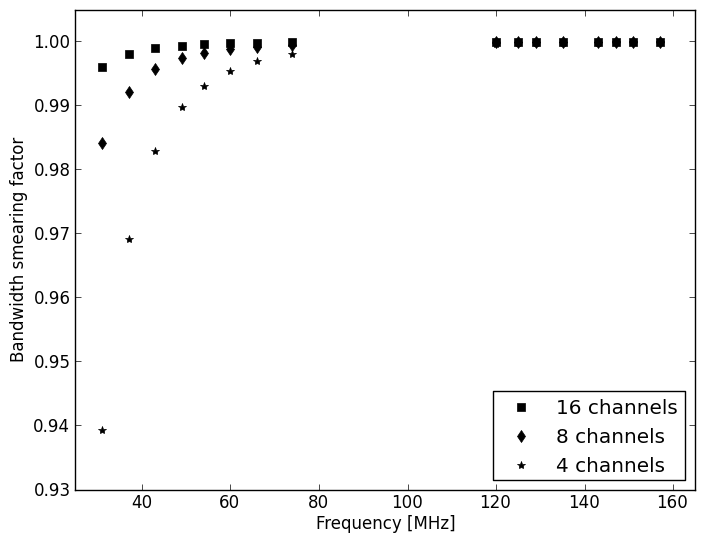}\\
\includegraphics[width=\hsize]{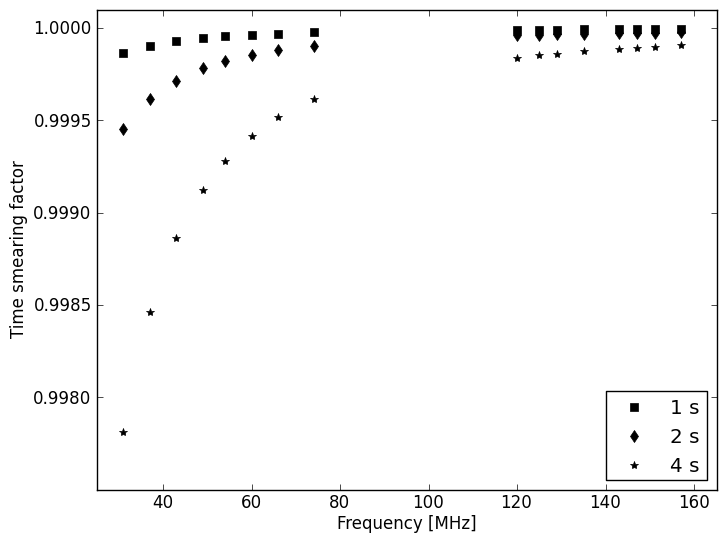}
\caption{Bandwidth ({\it top}) and time ({\it bottom}) smearing factors relevant to LOFAR MSSS observations, expressed as percentages and where unity is equivalent to no smearing. These values are calculated for $2\arcmin$ angular resolution and for sources at the half-power point of the station beam. In the top panel, points are plotted for averaging single subbands to 4 ({\it stars}), 8 ({\it diamonds}), and 16 ({\it squares}) channels. In the bottom panel, the smearing is calculated for visibility time averaging intervals of 1 ({\it squares}), 2 ({\it diamonds}), and 4 ({\it stars}) seconds.}
\label{figure:smearing}
\end{figure}

For LOFAR, the individual subband width (using the standard 200\,MHz clock) is 195.3125~kHz and in MSSS observations is divided into 64 channels. For a characteristic angular resolution of $2\arcmin$, we have calculated smearing factors for each of the MSSS band frequencies, and a field radius corresponding to half of the station beam HPBW (see \S\,\ref{subsection:lbasetup} and \ref{subsection:hbasetup}). The resulting bandwidth smearing curves have been calculated for different frequency averaging parameters and are displayed in Fig.~\ref{figure:smearing}. 

In the case of the LBA survey, at least 8 channels per subband must be retained in order to keep the effect of smearing to $\lesssim$1--2\%. For the HBA portion, the frequency averaging does not play such a crucial role. We thus retain 8 channels per subband in LBA for a maximum allowable bandwidth smearing factor at the lowest frequencies. In HBA, we retain 4 channels per subband to allow reprocessing at higher angular resolution without unacceptable smearing losses. Note that in the LBA, reprocessing the data in order to image at higher resolution will require either redoing the pre-processing steps with a lower frequency averaging factor, or acceptance of a substantial smearing factor at the lowest frequencies.

\subsubsection{Temporal resolution}

Another effect that needs to be accounted for is time-average smearing. Again referring to the expressions given by \citet{bridle_schwab_1999}, we assess the impact on MSSS data using:
\begin{equation}
\frac{I}{I_0}=1-1.22\times10^{-9}\left(\frac{r}{\theta}\right)^2\tau_\mathrm{a}^2
\end{equation}
where $\tau_\mathrm{a}$ is the averaging time. The estimated time smearing factors are shown in Figure \ref{figure:smearing}.

The effect of time smearing is considerably less than the impact of bandwidth smearing. For small time averaging intervals, the effect is negligible. The smearing is also less important than the need to retain high time resolution for recovery of ionospheric phase disturbances during the calibration process. We average the time steps to 2 seconds in order to recover high-quality station gain phases. This is illustrated in \S\,\ref{subsection:tip}.

\subsection{Primary and secondary calibration}

The primary calibrators for MSSS are listed in Table \ref{table:calibrators}, and are based on the calibration model presented by \citet{scaife_heald_2012}. The tabulated spectral coefficients correspond to that formulation, namely the $A_n$ ($n\geq1$) factors in
\begin{equation}
\log{S_{\nu}(\nu)} = \log A_0 + A_1\,\log \nu + A_2\,\log^2 \nu + \ldots
\end{equation}
These sources are mostly chosen to be compact on Dutch baselines (approximately 80 km or less), bright enough to give sufficient signal-to-noise per visibility, and have well-characterized radio spectra. The exception is Cygnus A, which is used as a primary calibrator in the LBA portion of MSSS, despite being a very complicated extended source. It does however have a very well determined source model, based on extensive commissioning work \citep[summarized by][]{mckean_etal_2011,mckean_etal_2015}.

\begin{table*}
\caption{Primary MSSS calibrator sources}
\label{table:calibrators}
\centering
\begin{tabular}{llllllll}
\hline\hline
Source ID & RA (J2000.0) & Dec (J2000.0) & $S_{150\,\mathrm{MHz}}$ (Jy) & Spectral coefficients & Morphology & LBA calibrator \\
\hline
3C48    & 01$^\mathrm{h}$37$^\mathrm{m}$41\fs3 & +33\degr09\arcmin35\arcsec & 64.768  & (-0.387,-0.420,0.181)        & point & No \\
3C147   & 05$^\mathrm{h}$42$^\mathrm{m}$36\fs1 & +49\degr51\arcmin07\arcsec & 66.738  & (-0.022,-1.012,0.549)        & point & No \\
3C196   & 08$^\mathrm{h}$13$^\mathrm{m}$36\fs0 & +48\degr13\arcmin03\arcsec & 83.084  & (-0.699,-0.110)              & double & Yes \\
3C286   & 13$^\mathrm{h}$31$^\mathrm{m}$08\fs3 & +30\degr30\arcmin33\arcsec & 27.477  & (-0.158,0.032,-0.180)        & point & No \\
3C295   & 14$^\mathrm{h}$11$^\mathrm{m}$20\fs5 & +52\degr12\arcmin10\arcsec & 97.763  & (-0.582,-0.298,0.583,-0.363) & double & Yes \\
3C380   & 18$^\mathrm{h}$29$^\mathrm{m}$31\fs8 & +48\degr44\arcmin46\arcsec & 77.352  & (-0.767)                     & point+diffuse & No \\
CygA    & 19$^\mathrm{h}$59$^\mathrm{m}$28\fs3 & +40\degr44\arcmin02\arcsec & 10690.0 & (-0.670,-0.240,0.021)        & FRII & Yes \\
\hline
\end{tabular}
\end{table*}

Observational verification of these primary calibrator sources started early in the MSSS test program, and revealed that while the brightest sources in the low band (3C196, 3C295, and CygA) were suitable primary calibrators, the others were too weak and/or too close to A-team sources to provide stable gain amplitude solutions. These fainter sources are still useful for the HBA portion of the survey but are not utilized in the LBA portion.

Primary (flux) calibration is handled with two different strategies in the LBA and HBA parts of the survey. In the LBA part, we take advantage of the fact that the individual dipoles are sensitive to emission from the entire visible sky, and there is no analog beamformer limiting the field of view. This allows us to observe with a simultaneous calibrator beam. The calibrator at the highest elevation angle is used, regardless of the distance between calibrator and target fields. The calibrator beam uses the exact same frequency coverage as the target fields and runs for the full length (11 minutes) of the target snapshots. This ensures sufficient time samples to obtain a stable gain amplitude solution. The CP obtains, and then exports, the median gain amplitude solution per station, per snapshot (along with the corresponding gain phase that is not used) from the calibrator beam, for application to the flagged and demixed target data in the TPP. On the other hand, a different strategy is used for the HBA observations. The analog HBA tile beam limits the field of view to typically 23 degrees HPBW at 150 MHz. This means that a bright, compact calibrator is not always available within the field of view near the targets. Therefore, the calibrator is observed alone (before the target snapshot) for 1 minute. Because the instantaneous sensitivity of the HBA system is much higher than that of the LBA system, stable gain amplitudes are obtained with a much shorter observing interval. As with LBA, the HBA calibrator beam covers the same frequencies as the target beams, and the CP exports the median gain amplitude per station. These are subsequently applied to the target snapshot data in the TPP.

Following the application of the primary flux calibration, the station phases remain uncalibrated (in the direction of the target field). The phase calibration takes place in the Target Imaging Pipeline, described in \S\,\ref{subsection:tip}.

\subsection{Automatic data quality filtering}

Much of the MSSS-LBA data were obtained early in LOFAR's lifetime, and this meant that much of the data were taken in unstable conditions. It turned out that typical observations included one or more bad stations (because of bad digital beam forming, network connection issues, or other reasons). We therefore incorporated conservative filtering steps into the pipeline, to identify and flag stations performing well outside of the normal bounds. The most important step considers the statistics of each station, and flags those that have an exceptionally large number of baselines with high measured noise. Before LOFAR's digital beams were well controlled, this step primarily removed stations with poorly focused beam responses.

\subsection{Target Imaging Pipeline}\label{subsection:tip}

The TIP stage of the MSSS pipeline combines the flagged, demixed, and flux-calibrated target snapshot observations to generate an initial image of the field. The TIP also optionally performs a self-calibration major cycle. Because it is decoupled from the pre-processing part of the pipeline, it can be run in an asynchronous manner with respect to the observational part of MSSS.

\begin{figure*}
\centering
\includegraphics[width=0.45\hsize]{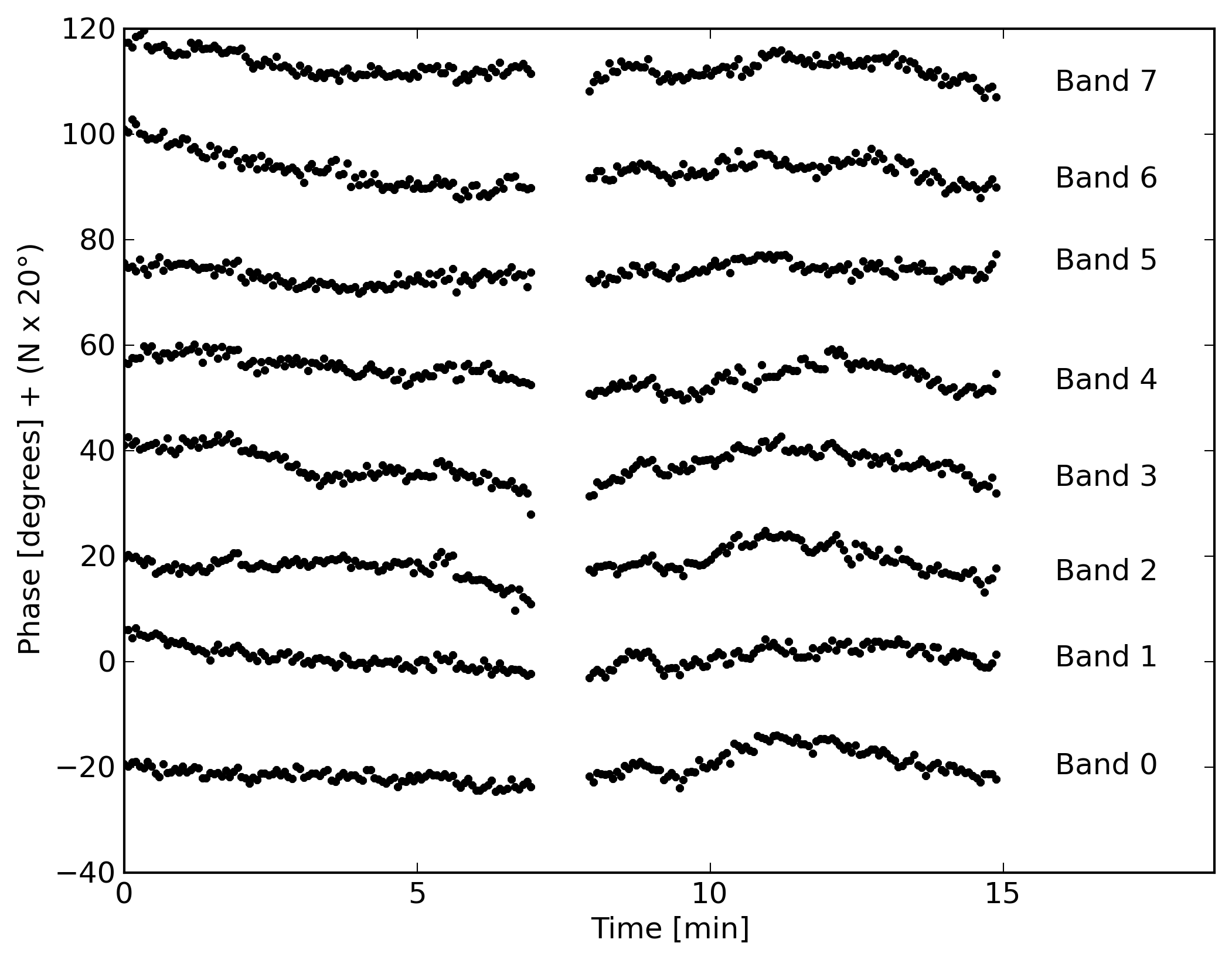}\,\includegraphics[width=0.45\hsize]{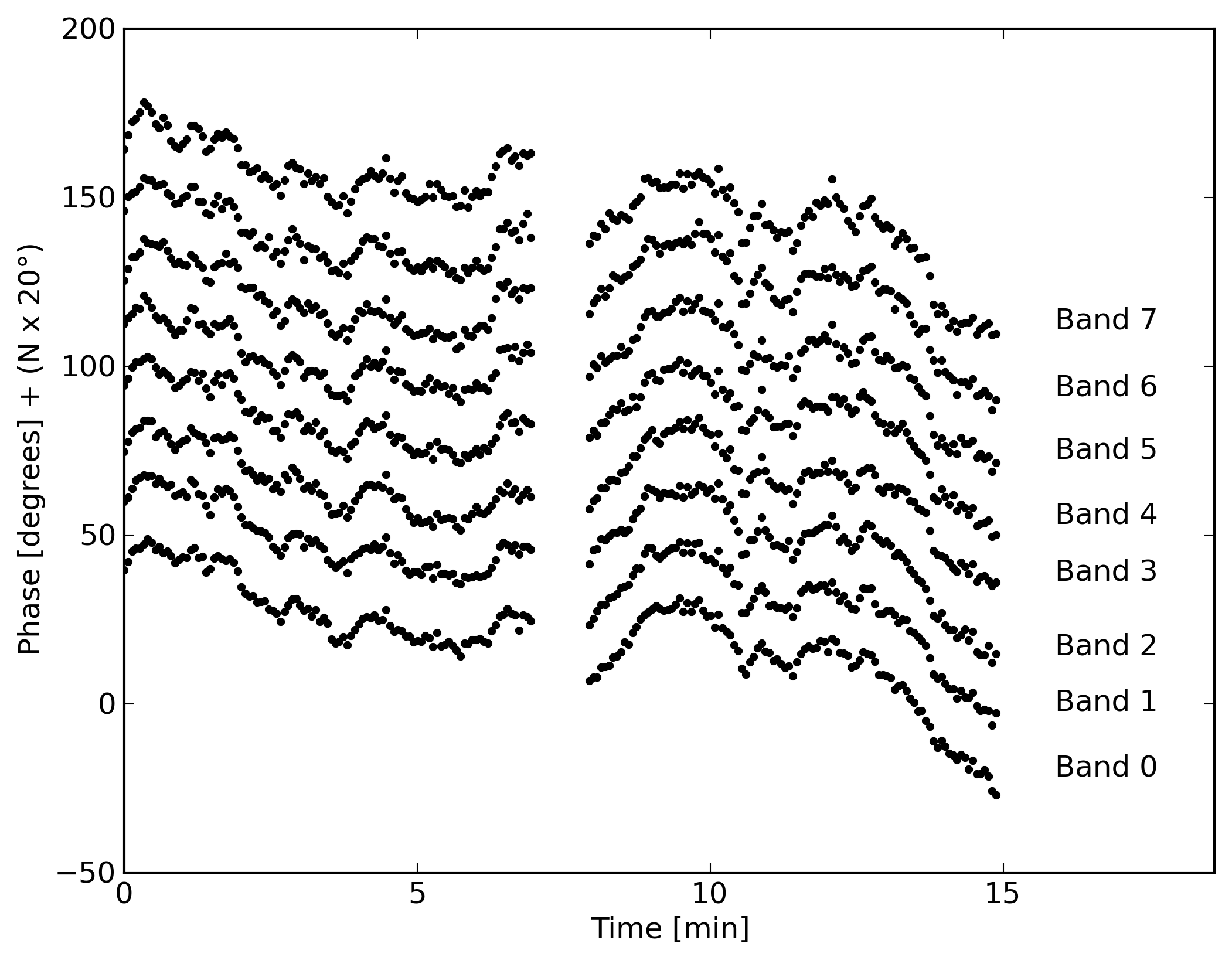}
\includegraphics[width=0.9\hsize]{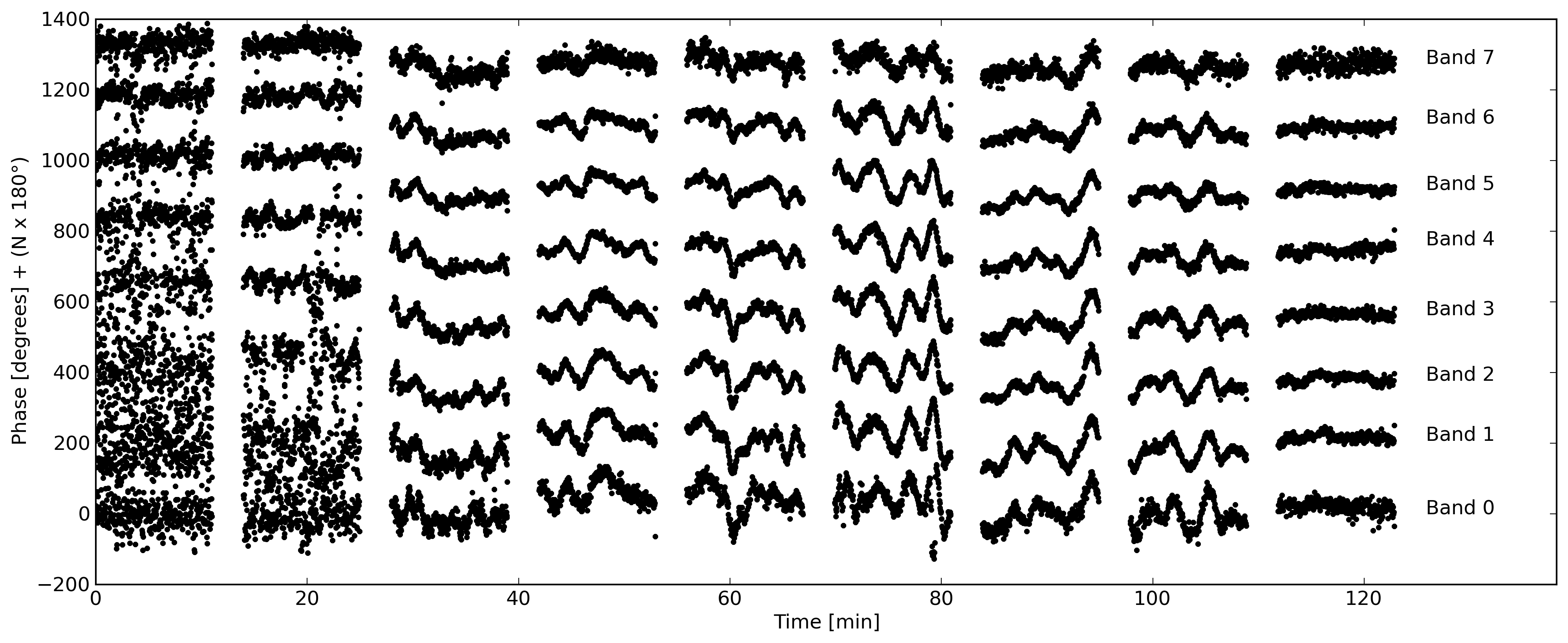}\\
\includegraphics[width=0.9\hsize]{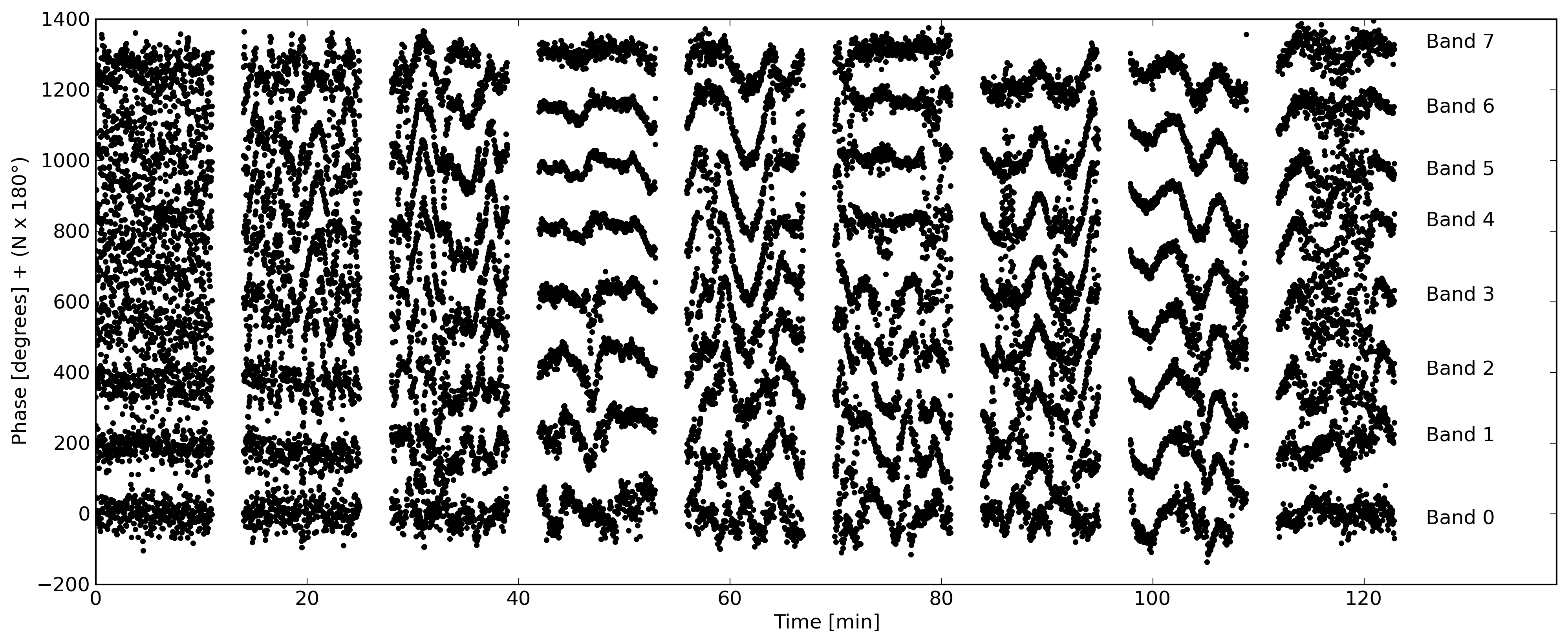}
\caption{Gain phases determined by the TIP (see \S\,\ref{subsection:tip}) for observations of the HBA and LBA fields H229+70 and L227+69 (see Table\,\ref{table:obslist}). Gain phases are displayed for HBA ({\it top row of panels}) and LBA ({\it bottom two panels}), referenced to CS002HBA0 and CS002LBA, respectively. (CS002 is one of the six superterp stations.) The phases are shifted vertically for display purposes. They correspond to the (0,0) element of the station gain matrix for CS302 ({\it top left and middle}) and RS306 ({\it top right and bottom}). The gaps between snapshots are compressed for display purposes, and correspond to 4 hours for HBA and 45 minutes for LBA.}
\label{figure:gainphases}
\end{figure*}

As a first step, the individual 2\,MHz band snapshots are combined, resulting in 8 visibility data sets per LBA or HBA field. These bands are treated separately throughout the TIP. The phase-only, direction-independent calibration is achieved using a VLSS-based sky model but taking the station sensitivity patterns into account. Example gain phases are shown in Fig.~\ref{figure:gainphases}. The phases are shown for two representative stations, CS302 (about 2 km southwest of the central group of six stations, collectively called the ``superterp'') and RS306 (about 15 km west of the superterp). These are shown as a function of time, with one set of phases for each of the 8 frequency bands, and displayed here with an arbitrary offset for visual clarity. Phase calibration is performed such that an independent solution is produced for each $2\,\mathrm{second}$ timestep.

\subsubsection{Imaging MSSS data}\label{subsection:imaging}

Imaging MSSS data is performed with the LOFAR imager, called the {\tt awimager} \citep{tasse_etal_2013}. It is based on the $\textsc{casa}$ imager including $w$-projection \citep{cornwell_etal_2008}, and also includes a LOFAR-specific implementation of A-projection \citep{bhatnagar_etal_2008} that treats the dipole and station response as time- and direction-dependent, full polarization terms in the measurement equation \citep[e.g.,][]{hamaker_etal_1996} during imaging and deconvolution. See \citet{tasse_etal_2013} for the results of simulations demonstrating the fidelity of the imaging step.

Our implementation of the LOFAR beam includes three layers. First, the LOFAR element beam is modelled through a full electromagnetic (EM) simulation (not including mutual coupling) and implemented in our software as a polynomial fit in elevation, azimuth, and frequency. The HBA analog tile beam is also included as the direct Fourier transform (DFT) of the dipole positions within a tile, and rotated to account for the orientation of each station \citep[as described by][the layout of the dipoles in each HBA station is rotated to reduce the sensitivity to bright far off-axis sources]{vanhaarlem_etal_2013}. Finally, the digital station beam for both LBA and HBA is calculated using a DFT of the dipole (LBA) or tile (HBA) positions in each individual station. Missing dipoles (LBA) and tiles (HBA) are indicated as such in the visibility data sets recorded by the correlator, and left out of the beam prediction during calculation. The digital station beams have been observationally mapped using the procedure described by \citet{vanhaarlem_etal_2013} and found to be in qualitative agreement with the predictions of the beam model. For MSSS data, the beam is applied in the {\tt awimager} such that it is considered to be constant within frequency blocks of width $100\,\mathrm{kHz}$, and time blocks of $5\,\mathrm{min}$.

Our initial imaging run per field incorporates projected baselines shorter than $2\,\mathrm{k}\lambda$. For fields at declination $\delta\leq35^\circ$, we leave out baselines shorter than $100\,\lambda$, which we found empirically to provide a smoother background. This imaging run is performed with a simple, shallow deconvolution strategy (using 2500 {\tt CLEAN} iterations). After recalibrating the data on the basis of this first imaging round (see \S\,\ref{subsection:recal}), we update the imaging parameters in preparation for final catalog creation: the maximum baseline length is increased to $3\,\mathrm{k}\lambda$. After an initial shallow deconvolution, we create a mask on the basis of a source detection round performed in the way described in \S\,\ref{section:sourcefinding}, and subsequently perform a deep deconvolution using the mask to limit the locations of {\tt CLEAN} components. The final deconvolution is limited by reaching a cutoff of $0.5\,\sigma$ instead of by limiting the number of components. All imaging steps use Briggs weighting \citep{briggs_1995} with a robust parameter of 0. Final image products are produced by mosaicing individual pointings in each band using the standard inverse variance weighting technique and making use of the predicted effective primary beam images that are produced as a standard output of the {\tt awimager}.

We note that we expect a low impact of {\tt CLEAN} bias \citep[i.e., the reduction in recovered flux density of real sources due to deconvolution of sidelobes inadvertently identified as true sources; e.g.][]{becker_etal_1995,cohen_etal_2007}. First, the snapshot $uv$ coverage is excellent for imaging at the low angular resolution that we make use of in MSSS (see Figures~\ref{figure:uvcoverage_lba} and \ref{figure:uvcoverage_hba}), meaning that the synthesized beam has low sidelobe levels (rms sidelobe levels of $\lesssim1\%$ in both HBA and LBA, with isolated maxima of $12-13\%$ in the HBA and isolated maxima very close to the main lobe of $16-28\%$ in the LBA). Second, the masked deconvolution that we employ ensures that we only apply {\tt CLEAN} to real sources. Since sidelobes shift position at different frequencies, our multi-frequency source detection mitigates the impact of sidelobes in any individual band. We characterize the {\tt CLEAN} bias present in MSSS-HBA data in \S\,\ref{subsection:cleanbias}.

A unique benefit of incorporating a time-, frequency-, and direction-dependent term in the imaging and deconvolution step through the A-projection algorithm is the ability to directly incorporate ionospheric corrections. We implement this procedure for our LBA data, as described in \S\,\ref{subsection:recal}. For the application of the ionospheric correction we increase the time resolution to $10\,\mathrm{s}$ in order to capture the rapidly variable phases.

\subsubsection{Source finding}\label{section:sourcefinding}

The source finding is performed using two complementary source extraction software packages: PyBDSM\footnote{{\tt http://tinyurl.com/PyBDSM-doc}} and PySE\footnote{{\tt http://docs.transientskp.org/tkp/master/tools/pyse.html}}. 

Both packages allow calculation of rms and mean images and identification of sources in radio maps through the use either of a False Detection Rate (FDR) method \citep{hopkins_etal_2002}, or of a threshold technique that locates islands of emission above some multiple of the noise in the image. Gaussians are then fitted to each island for accurate measurement of source properties. Both PyBDSM and PySE allow the simultaneous use of separate detection and analysis image files for island definition and Gaussian fitting respectively. We make use of this functionality to produce a reliable catalog that retains the full multifrequency information provided by the data.

While PySE has been designed as a source finding tool intended primarily for use by the LOFAR Transients Key Project and is therefore conceived for the detection of unresolved sources, PyBDSM has been programmed for the more general search of both compact and diffuse sources. PyBDSM thus allows fitting of multiple Gaussians to each island, or grouping of nearby Gaussians within an island into sources. In addition a PyBDSM module is available to decompose the residual image resulting from the normal fitting of Gaussians into wavelet images of various scales. This step is useful for automatic detection of diffuse sources. In the following, we therefore combine PyBDSM and PySE results only in the case of point-like sources. Our source finding strategy consists of running both PyBDSM and PySE separately and then combining the results as described in \S\,\ref{sec:cat}. Eight different joint PyBDSM and PySE catalogs are therefore initially produced for each of the 8-band maps in both HBA and LBA, and these are subsequently merged to give the final multi-frequency catalog.

We define two thresholds for the islands: one to determine the region within which source fitting is done, and another such that only islands with peaks above the threshold are used. For MSSS images, we set these two thresholds to $5\sigma$ and $7\sigma$, respectively. In addition, similarly to what is extensively done in the visible domain \citep[see e.g.][]{szalay_etal_1999}, we use an 8-band combined image (see below for a description) and each single band image as detection and analysis images respectively. The use of a combined image for island definition optimizes sensitivity to faint sources. Since the significant islands are identified using a single image, this procedure also alleviates the task of matching the eight single-band catalogs. Note however that the Gaussian fitting is performed on each single-band image independently. This results in possible differences between the central position of each source as a function of frequency. Therefore the resulting multi-band catalogs need to be matched as described in \S\,\ref{sec:cat}.

The 8-band mosaic images on which we run the source finder tools are produced using the same $uv$-range and convolved to a common resolution using the $\textsc{miriad}$ \citep{sault_etal_1995} task {\small CONVOL}. We do this both for primary-beam (PB) and non-primary-beam (NPB) corrected images. The latter are used to produce the combined mosaic image, obtained by performing an inverse-variance weighted average of the 8-band NPB corrected mosaic images (i.e., the weight per image is $w_i=1/\sigma_i^2$ where $\sigma_i$ is the rms of image $i$). This combined image is used as the detection image, while the individual 8-band PB corrected images are used as analysis images. In this way, we avoid fake detections in the image borders (which may be caused by the increase of the rms related to PB correction), while obtaining properly corrected flux densities per source in the output of the source finders.

For the combined catalog description and additional detailed information regarding the method by which it is produced, see \S\,\ref{sec:cat}.

\subsubsection{Calibration stability and ionospheric corrections}\label{subsection:recal}

To produce the final survey output, we run the TIP twice. The first time incorporates the VLSS-based sky model as described in \S\,\ref{subsection:tip}. For the second pass, we repeat the phase calibration using a sky model created from the first-pass calibration and imaging round. On the basis of this new phase calibration, we generate a new set of images (as described in \S\,\ref{subsection:imaging}) and source catalogs. The intention of this step is to minimize the effect of any spurious sources that may be present in the initial sky model and to ensure that the MSSS catalog is based on an internally consistent calibration cycle. This procedure has been followed for the representative data set considered in \S\,\ref{section:mvf}.

For the low frequency and large field of view intrinsic to LOFAR observations in the LBA band, ionospheric effects are strong even when imaging at the modest $\approx2\arcmin$ resolution utilized for MSSS. The clearest effect in the image plane is the presence of spiky artefacts around the brightest sources. To deal with this we have implemented a scheme similar to the Source Peeling and Atmospheric Modeling \citep[SPAM;][]{intema_etal_2009} approach that has been successfully used for VLA and GMRT data. We briefly summarize the procedure here; a full description will be provided in a forthcoming publication. The sky model is divided into approximately 30 source groups, and each group is used to derive a phase solution at each frequency. These 30 source groups cover the entire field of view visible at 30~MHz. To trace the frequency behavior of the calibration phases in those directions at the highest frequencies, where the field of view is much smaller, we utilize the simultaneously observed flanking beams. Thus ionospheric phases are only available across the full field of view for the central field in a multiplexed observation like MSSS. For the dataset considered in \S\,\ref{section:mvf}, only the central field (L227+69) can be processed in this manner.

The frequency dependent phases include two terms: a non-dispersive clock delay term\footnote{LOFAR stations only share a single clock within the core area; the remote stations are on independent clocks.} and a dispersive ionospheric delay term. The clock term is constant in all directions at a given time, while the dispersive term is direction-dependent. To isolate the ionospheric term we subtract the phases determined in one direction from those in all other directions, leaving a differential ionospheric phase term per direction and per station. By considering the phase of each station in each direction as a single ``pierce point'' through a single thin-layer ionosphere, these values are used to fit a ``screen'' of ionospheric total electron content (TEC) at a particular height. The difference between data and fitted screen is used to identify an optimized TEC screen height, typically around 100--200\,km. The TEC screen can be used as an input to the {\tt awimager} to correct the phase distortions across the field of view, as a function of both frequency and time.

\section{Standard data products}\label{section:products}

The primary output of MSSS is a catalog containing positions and Stokes I flux densities for all confidently detected sources, as well as extents and orientations for resolved sources. Spectral behavior of the sources is also provided. The MSSS catalog is stored in the LOFAR GSM database, implemented in a fully VO-compliant system based on the Data Center Helper Suite \citep[DaCHS;][]{demleitner_2014}.

Processed visibilities will be stored in the LOFAR long term archive (LTA) for future reprocessing, for example by science groups that wish to reprocess the data to search for polarization, or to perform long-baseline imaging (see \S\,\ref{section:science}). In addition, raw visibilities are always stored in the LTA immediately after observation. At the conclusion of the TIP, the images are imported to a postage-stamp server\footnote{MSSS data are being hosted at {\tt http://vo.astron.nl} where the subset presented in \S\,\ref{section:mvf} has already been made available.}, which allows inspection of images with catalog overlays and multi-frequency source spectrum pop-up plots, as well as providing direct-download links to FITS files.

Following verification of the survey results, the survey output (images and catalogs) will become fully public.

\subsection{MSSS Images}

The MSSS image products will be released in {\tt FITS} format, and will consist of mosaics corresponding to the fields specified in \S\,\ref{subsection:mosaicgrid}. Each mosaic consists of sixteen 2~MHz bandwidth images at the central frequencies listed in Table \ref{table:surveyconfig}, and two 16~MHz full bandwidth images, one for LBA and the other for HBA.

\subsection{MSSS Catalog}\label{sec:cat}

The MSSS catalog is generated in several steps. First, the PySE and PyBDSM source finders are run on all individual HBA and LBA frequency bands using the combined maps as detection images (see Section \ref{section:sourcefinding} for details). When deriving source positions, sizes and fluxes, both PyBDSM and PySE take into account fitting errors caused by correlated map noise following the approach suggested by \citet{condon_1997}. Ionospheric phase calibration is the other largest contributor to the positional uncertainty of fitted sources \citep{cohen_etal_2007}. We describe in the following how these errors have been taken into account when producing the multi-band final catalog.

We firstly work on the HBA and LBA catalogs separately. For each source finder, the detected sources are associated across the eight individual frequency bands in order to generate a concatenated multi-frequency source catalog. The association is performed by calculating the angular distance between sources in any pair of individual frequency band catalogs. A match is determined to be positive if each element of the pair is mutually the nearest to its counterpart from the other catalog. This criterion ensures that sources have exclusive pairing, as opposed to possible multiple associations. An additional threshold is applied in order to reject sources that are too distant and might be spuriously associated; hence we require ${\rm distance} \leq 3 \sqrt{\sigma_1^2 + \sigma_2^2}$, where $\sigma_{1,2}$ are the fitted positional uncertainties in the first and second catalog, respectively. Note that, in this step, we do not take into account calibration related errors, since in each of the two segments (HBA and LBA) the 8 bands have been observed simultaneously. We chose a threshold of $3\sigma$ after verifying that the curve of growth of positive matches started plateauing around this level. After the source association is completed, we calculate the position of each source -- RA and Dec separately -- as the weighted average position among the frequency bands in which it was detected taking into account the positional uncertainties. The uncertainties associated with the average positions are calculated by propagating the errors accordingly.

Following the source association that has so far been performed only within the PySE and the PyBDSM catalogs, sources detected with each source finder are also cross-matched in order to generate the final catalog. The same pairing process as explained above is used for this step. However, in order to preserve consistency in the final catalog, we did not attempt to calculate average values for the various reported fields. Rather, we chose the PyBDSM fields to prevail over those from PySE when both values existed. Hence, the reported IDs and positions are taken from PyBDSM unless a source was only detected by PySE. Similarly, for a given frequency band, the reported flux density properties are those found by PyBDSM unless the source was only detected in this particular frequency band by PySE. Each frequency band possesses a field {\tt SFFLAG{\it nnn}} that indicates which, if not none or both, source finder the source was detected with at frequency {\tt {\it nnn}} MHz. By considering the results from both source finders together in this way, we can add confidence to the reality of individual detections; thus we recommend using sources detected by only one source finder with caution.

In order to produce the final source list and associate the HBA and LBA catalogs, we need to incorporate an estimate of the effect on source positions caused by calibration errors. Following \citet{cohen_etal_2007}, we compare our HBA and LBA catalogs to the NRAO/VLA Sky Survey \citep[NVSS;][]{condon_etal_1998} catalog, which has higher resolution and S/N, as well as significantly lower calibration errors due to the higher observing frequency (1.4~GHz). At each of the two frequencies we identify sources with i) a detection level of at least $30 \sigma$, ii) a single bright\footnote{Peak flux higher than 50 mJy beam$^{-1}$.} NVSS counterpart within 1.5 the beamsize, and iii) a fitted major axis less than 1.5 the beamsize. For example, in the case of the LBA data presented in \S\,\ref{section:mvf}, we derive an average offset of $\Delta{\rm RA}_{\rm mean}=0\farcs18$ and $\Delta{\rm Dec}_{\rm mean}=0\farcs03$ with associated rms deviations from the mean values of $\Delta{\rm RA}_{\rm rms}=1\farcs59$ and $\Delta{\rm Dec}_{\rm rms}=0\farcs24$. For the HBA data in \S\,\ref{section:mvf}, where no direction dependent calibration was applied, the coordinates display larger offsets: $\Delta{\rm RA}_{\rm mean}=2\farcs18$ and $\Delta{\rm Dec}_{\rm mean}=-0\farcs79$ with rms deviations from the mean values of $\Delta{\rm RA}_{\rm rms}=2\farcs92$ and $\Delta{\rm Dec}_{\rm rms}=2\farcs45$. During the production of the final multi-band (16-frequency) catalog, these calibration errors are added in quadrature to the positional uncertainties of the fitted sources. The HBA and LBA association is subsequently performed as described before, but taking into account both fitting and calibration position errors.

Finally, we perform a post-concatenation analysis in order to determine the spectral properties for each source. For this step, only the PyBDSM fluxes are used -- again this is done in order to ensure consistency since the flux scale between the two source finders may suffer from biases. The spectrum of each source is fitted with the functional form \citep[see also][]{scaife_heald_2012}:
\begin{equation}\label{eqn:spix}
S_\nu(\nu)=A_0\,10^{A_1\log(\nu/150\,\mathrm{MHz})}.
\end{equation}
Given the large number of sources to be fitted and the fact that the posterior distribution of the spectral fit should be well-behaved -- this is a linear least-squares fit to a polynomial in $\log S_\nu$ space -- we use a Levenberg-Marquardt $\chi^2$ minimisation algorithm to determine the best-fit parameters and errors.

We use a locally determined effective beamsize to deconvolve the
source sizes reported by PyBDSM, and classify sources as extended if
the deconvolved size is nonzero. Since most sources are unresolved at this resolution, this procedure allows us to mitigate the effect of ionospheric smearing.

The 214 MSSS mosaics overlap at their edges to ensure that all sources are reliably imaged and cataloged. This means that many sources are identified in more than one mosaic. After creating a catalog from each mosaic as described above, we filter the multiply cataloged sources to remove duplicate entries. First, we look for sources that have the same source ID, which identifies those having the same coordinates at arcsecond precision in both RA and Dec. Since the mosaics are formed from the same images, most sources that are present in more than one mosaic are identified at the same coordinates. However, small differences sometimes arise because the local backgrounds and noise levels calculated by the source finders differ slightly in neighboring mosaics. Therefore, an extra sifting is performed by identifying the nearest neighbour of each cataloged source. Such neighbors are matched and removed if they were found in different mosaics, and if their separation is less that $45\arcsec$ (substantially smaller than the resolution, but large enough to identify matching source pairs even when their central position is less precise).

The MSSS catalog has the following columns. First, a set of parameters that are common for both the point source catalog and extended source catalog:
\begin{description}[style=multiline,leftmargin=2.5cm,font=\tt]
\item[ID] Source ID, formed as ``1MSSS J$hhmmss+ddmmss$'' using the IAU convention. In the catalog presented in \S\,\ref{section:mvf}, we instead use the source ID prefix ``MSSSVF'' to distinguish it from the forthcoming full MSSS catalog.
\item[RA] Source J2000 Right Ascension, in decimal degrees.
\item[DEC] Source J2000 Declination, in decimal degrees.
\item[e\_RA] Error in source J2000 Right Ascension, in seconds of time. This is a formal error based on the source position fit.
\item[e\_DEC] Error in source J2000 Declination, in arcseconds. This is a formal error based on the source position fit.
\item[e\_RA\_sys] Full error in source J2000 Right Ascension, in seconds of time. This includes both the formal error based on the source position fit and a systematic positional error term.
\item[e\_DEC\_sys] Full error in source J2000 Declination, in arcseconds. This includes both the formal error based on the source position fit and a systematic positional error term.
\item[SFFLAG{\it nnn}] Flag indicating which source finder identified the source at {\it nnn} MHz (0 means it was detected in both; 1 means it was detected only in PyBDSM; 2 means only in PySE; 3 means no detection). Note that if the source was identified by both source finders, the reported flux density values are those of PyBDSM.
\item[Sint{\it nnn}] Source integrated flux density at {\it nnn} MHz, in Jy.
\item[e\_Sint{\it nnn}] Error in source integrated flux density at {\it nnn} MHz, in Jy.
\item[Spk{\it nnn}] Source peak flux density at {\it nnn} MHz, in Jy\,beam$^{-1}$.
\item[e\_Spk{\it nnn}] Error in source peak flux density at {\it nnn} MHz, in Jy\,beam$^{-1}$.
\item[A0\_LBA] Spectral flux density at 150 MHz, A$_0$ in Equation \ref{eqn:spix}. Derived from LBA values only.
\item[e\_A0\_LBA] Error in spectral flux density at 150 MHz. Derived from LBA values only.
\item[A1\_LBA] Spectral index, A$_1$ in Equation \ref{eqn:spix}. Derived from LBA values only.
\item[e\_A1\_LBA] Error in spectral index. Derived from LBA values only.
\item[A0\_HBA] Spectral flux density at 150 MHz, A$_0$ in Equation \ref{eqn:spix}. Derived from HBA values only.
\item[e\_A0\_HBA] Error in spectral flux density at 150 MHz. Derived from HBA values only.
\item[A1\_HBA] Spectral index, A$_1$ in Equation \ref{eqn:spix}. Derived from HBA values only.
\item[e\_A1\_HBA] Error in spectral index. Derived from HBA values only.
\item[A0] Spectral flux density at 150 MHz, A$_0$ in Equation \ref{eqn:spix}.
\item[e\_A0] Error in spectral flux density at 150 MHz.
\item[A1] Spectral index, A$_1$ in Equation \ref{eqn:spix}.
\item[e\_A1] Error in spectral index.
\item[NDET] Number of bands in which the source was detected.
\item[NDET\_LBA] Number of LBA bands in which the source was detected.
\item[NDET\_HBA] Number of HBA bands in which the source was detected.
\item[NUNRES] Number of bands in which the source is unresolved.
\item[NUNRES\_LBA] Number of LBA bands in which the source is unresolved.
\item[NUNRES\_HBA] Number of HBA bands in which the source is unresolved.
\item[MOS\_ID] Mosaic name from which the source was extracted.
\item[CAL\_ID\_LBA] Primary flux calibrator(s) used to set the LBA flux density scale in the vicinity of the source.
\item[CAL\_ID\_HBA] Primary flux calibrator(s) used to set the HBA flux density scale in the vicinity of the source.
\end{description}
A set of parameters that are only included for the extended source catalog:
\begin{description}[style=multiline,leftmargin=2.5cm,font=\tt]
\item[MAJAX{\it nnn}] Major axis of fitted ellipse at {\it nnn} MHz, in decimal degrees.
\item[MINAX{\it nnn}] Minor axis of fitted ellipse at {\it nnn} MHz, in decimal degrees.
\item[PA{\it nnn}] Position angle of fitted ellipse at {\it nnn} MHz, in decimal degrees.
\item[e\_MAJAX{\it nnn}] Error in major axis, in decimal degrees.
\item[e\_MINAX{\it nnn}] Error in minor axis, in decimal degrees.
\item[e\_PA{\it nnn}] Error in position angle, in decimal degrees.
\end{description}

This catalog has 108 columns for the point sources, and $108+96=204$ columns for extended sources. Assuming that there will be $10^5$ sources (see \S\,\ref{section:mvf}) with about 5\% of those extended, then that will lead to about 11.3 million cataloged data values.

\section{Initial MSSS imaging results}\label{section:mvf}

MSSS observations are typically performed for several hours per week within the rest of the LOFAR schedule. The HBA segment has been fully observed and initial images have been created. While the survey calibration processing is still ongoing, we highlight a representative portion of the sky to illustrate the output that will be forthcoming for the full northern sky. For this we selected a 100 square degree patch of sky that has subsequently been used repeatedly to test imaging pipeline performance. The field was randomly selected and we refer to it here as the MSSS Verification Field (MVF). It includes a few moderately bright point-like sources but no bright, complicated 3C or 4C sources, and is otherwise distant from the troublesome A-team sources. The Galactic contribution to the sky brightness is relatively unimportant in this direction (Galactic coordinates $(l=108^\circ,b=44^\circ)$ at the center of the MVF).

\begin{figure*}
\includegraphics[width=0.5\hsize]{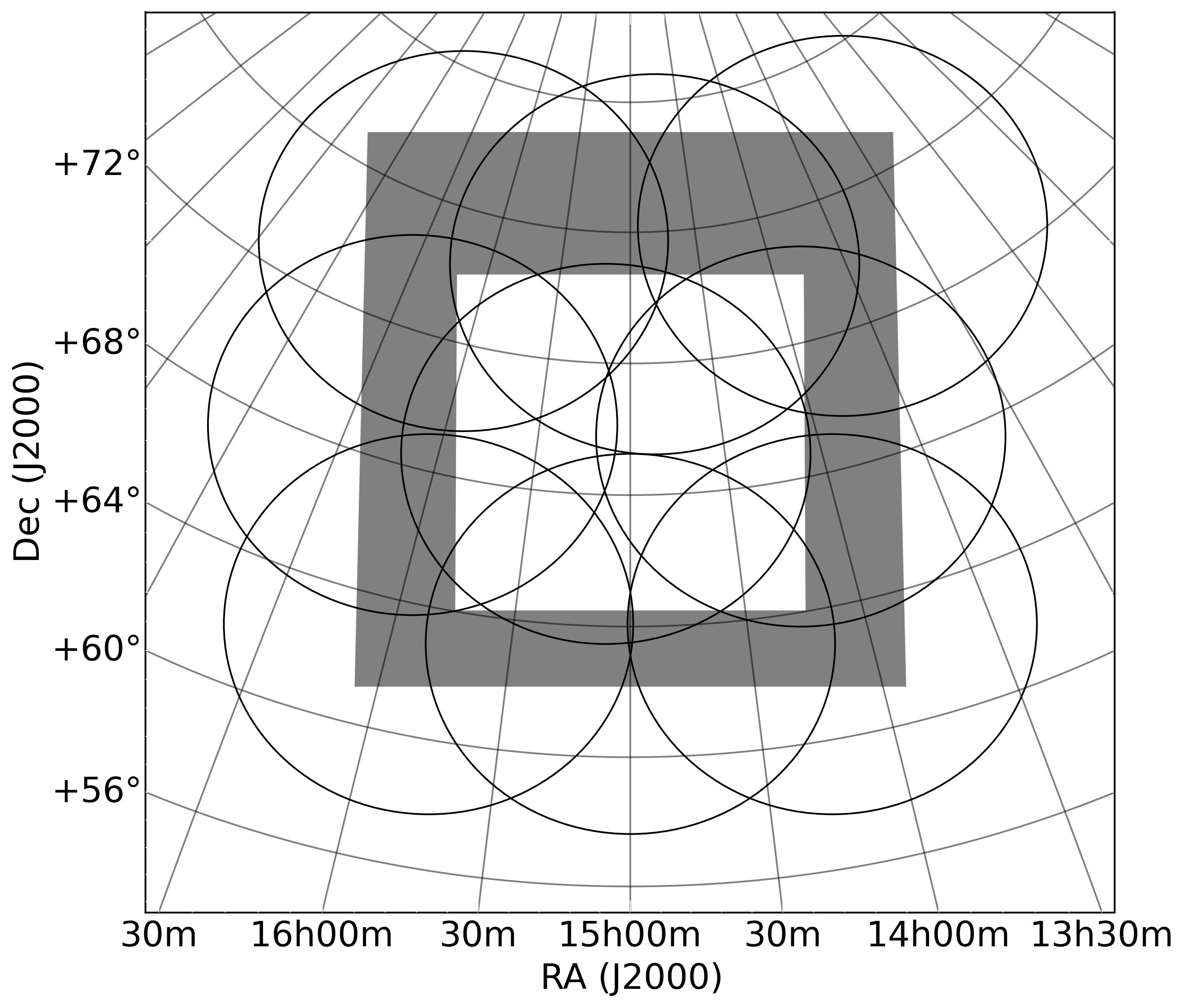}\includegraphics[width=0.5\hsize]{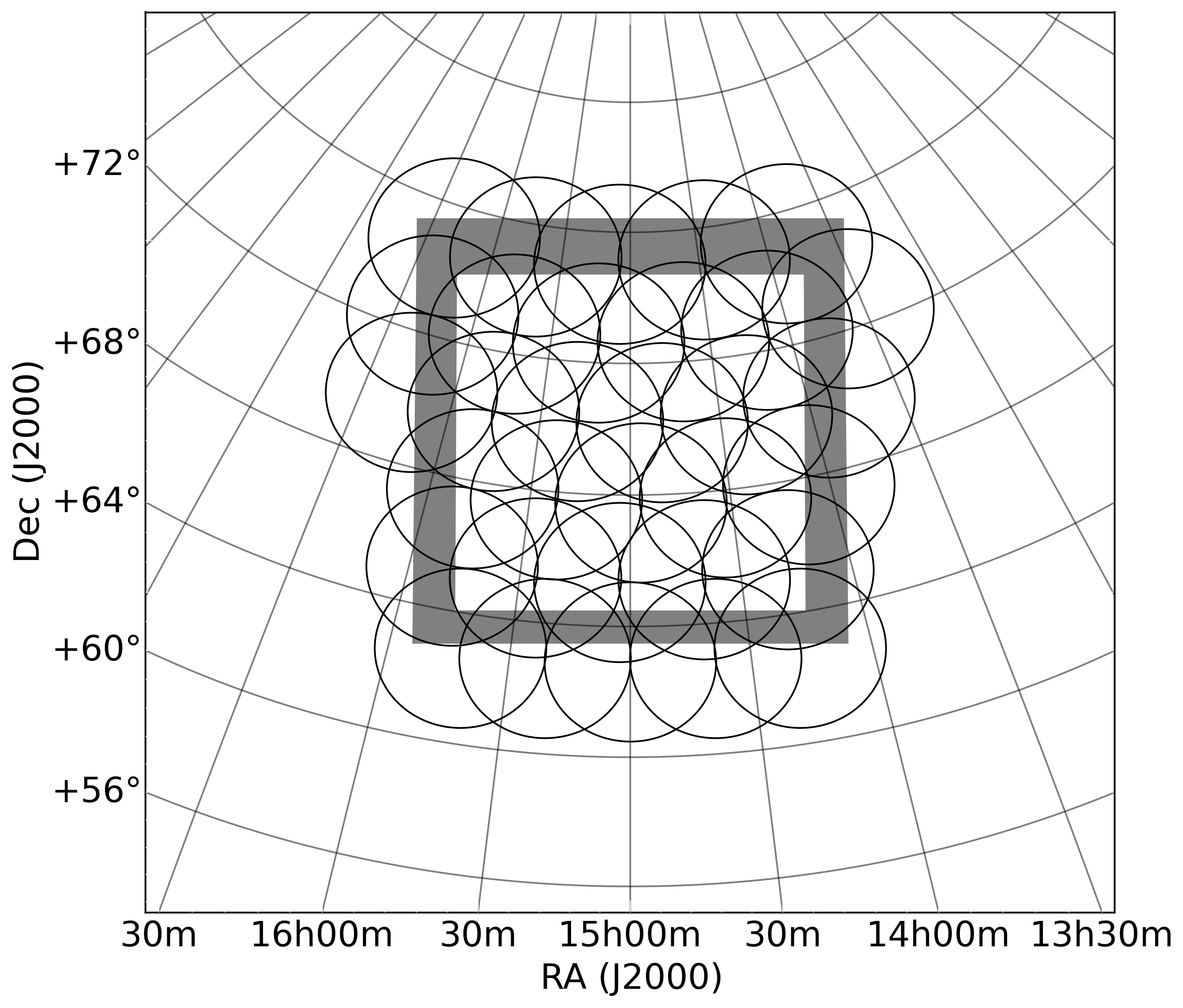}
\caption{Mosaic layout of the MSSS Verification Field (MVF). {\it Left}: The nine LBA fields making up the MVF, overlaid on the $10\degr\times10\degr$ mosaic field (central white square). The gray border around the mosaic area is a guard area used to ensure that sufficient edge fields are included and to ensure flat sensitivity within the mosaic. {\it Right}: The 32 HBA fields making up the MVF, overlaid on the same central mosaic field. The guard border is smaller in proportion to the HBA field diameter.}
\label{figure:mosaics}
\end{figure*}

\begin{figure*}
\centering
\includegraphics[width=0.8\hsize]{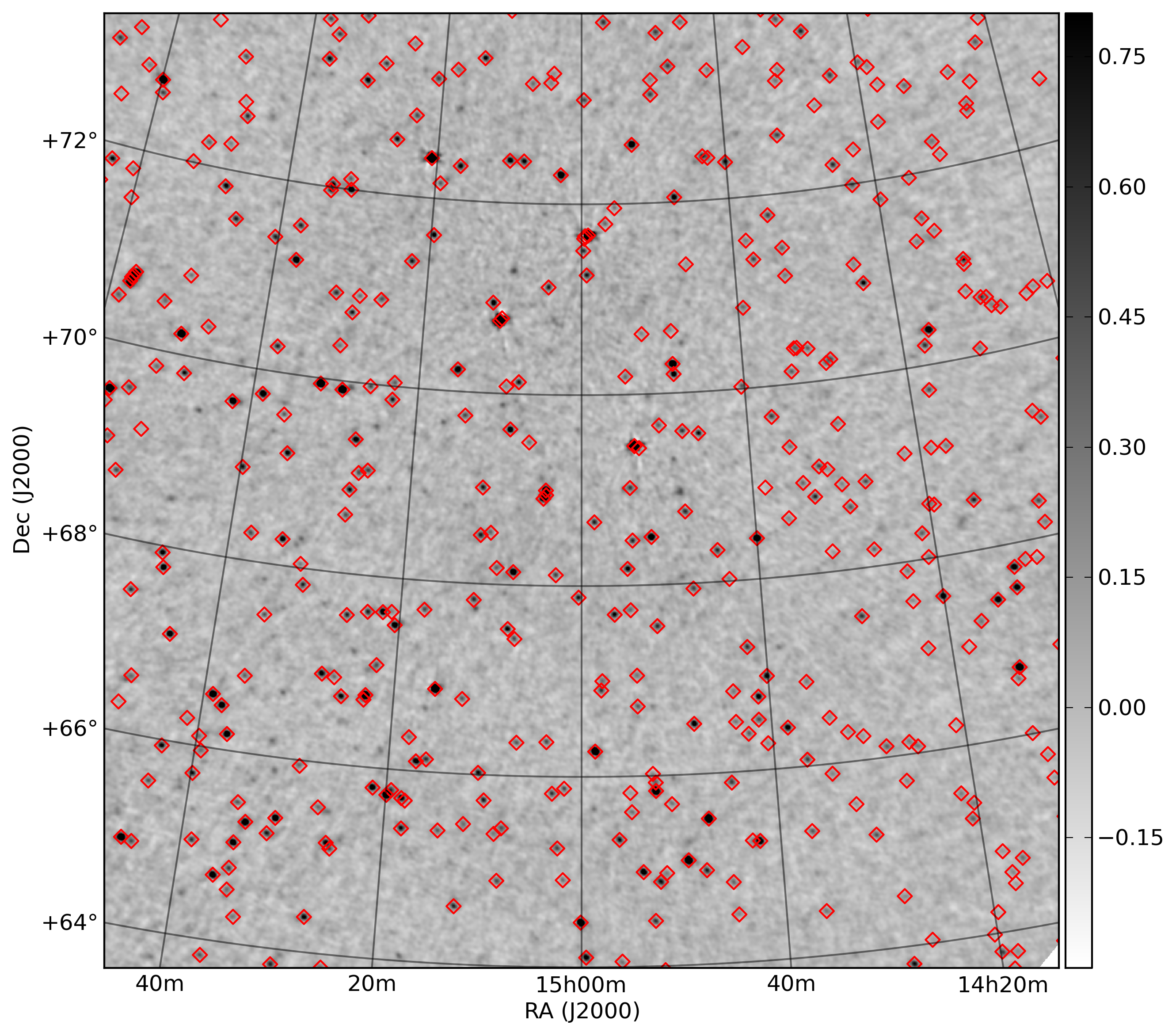}
\caption{LBA full-bandwidth central field of the MVF, after ionospheric correction and displayed here without primary beam correction for clarity. The noise level is $39\,\mathrm{mJy\,beam}^{-1}$, and the synthesized beam is $166\arcsec$. The colorbar units are $\mathrm{Jy\,beam}^{-1}$. Diamonds mark the positions of cataloged VLSSr sources.}
\label{figure:mvflbamosaic}
\end{figure*}

\begin{figure*}
\centering
\includegraphics[width=0.8\hsize]{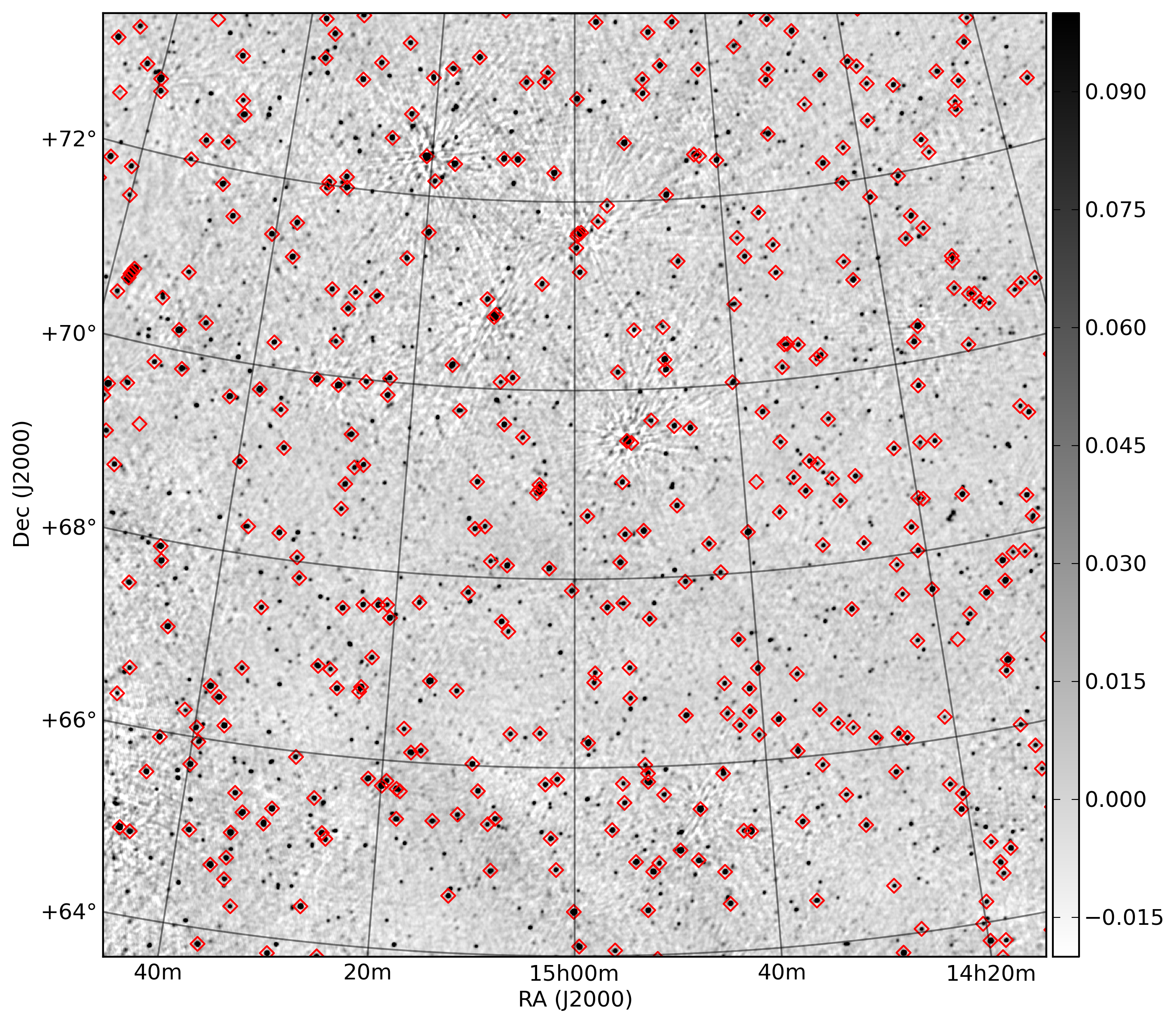}
\caption{HBA full-bandwidth mosaic of the MVF, displayed here without primary beam correction for clarity. The noise level is $5\,\mathrm{mJy\,beam}^{-1}$, and the synthesized beam is $108\arcsec$. The colorbar units are $\mathrm{Jy\,beam}^{-1}$. Diamonds mark the positions of cataloged VLSSr sources.}
\label{figure:mvfhbamosaic}
\end{figure*}

The mosaics in this region are created from 9 LBA fields and 32 HBA fields, as illustrated in Fig.~\ref{figure:mosaics}. These fields were not all observed at the same time or even on the same day. The observing summary is listed in Table \ref{table:obslist}.
The primary flux calibrator for fields H244+70 and H245+75 was 3C380, and the primary flux calibrator for all other fields was 3C295. After convolution to a common beam size, the effective resolution is $108\arcsec$ in HBA, and $166\arcsec$ in LBA. Images of the frequency-averaged LBA field and HBA mosaic are shown in Figs.~\ref{figure:mvflbamosaic} and \ref{figure:mvfhbamosaic}, respectively.

Due to the ionospheric processing scheme described in \S\,\ref{subsection:recal}, only the central LBA field (L227+69) is used to generate the LBA part of the MVF catalog. Moreover, only 7 of the 9 LBA snapshots were used due to particularly bad ionospheric quality during the first 2 snapshots (see Fig.~\ref{figure:gainphases}). The HBA portion of the catalog was produced based on the combination of all fields listed in Table~\ref{table:obslist}. All told, 1234 unique sources were identified within the 100 square degrees of the MVF (307 in LBA, 1234 in HBA). A simple projection to the full MSSS survey area would suggest that approximately 250\,000 sources will be found, but taking into account reduced sensitivity at low declination and Galactic latitude, as well as poor image quality near extremely bright sources, we expect to recover between 150\,000--200\,000 sources in the full MSSS catalog.

\begin{table*}
\caption{MSSS Verification Field Observing Log}
\label{table:obslist}
\centering
\begin{tabular}{ll|ll|ll|ll|ll}
\hline\hline
Field & Date & Field & Date & Field & Date & Field & Date & Field & Date \\
\hline
L212+63 & 2013 Apr 8  & H214+63 & 2013 Apr 21 & H231+65 & 2013 Apr 21 & H215+70 & 2013 Apr 21 & H228+73 & 2013 Feb 15 \\
L225+63 & 2013 Mar 18 & H220+63 & 2013 Apr 21 & H237+65 & 2013 Apr 21 & H222+70 & 2013 Feb 15 & H236+73 & 2013 Feb 15 \\
L238+63 & 2013 Apr 8  & H225+63 & 2013 Apr 21 & H212+68 & 2013 Apr 21 & H229+70 & 2013 Feb 15 & H244+73 & 2013 Apr 21 \\
L211+69 & 2013 Mar 18 & H230+63 & 2013 Apr 21 & H218+68 & 2013 Feb 15 & H236+70 & 2013 Feb 15 & H208+75 & 2013 Apr 21 \\
L227+69 & 2013 Mar 18 & H236+63 & 2013 Apr 21 & H224+68 & 2013 Feb 15 & H244+70 & 2013 Feb 22 & H217+75 & 2013 Apr 21 \\
L243+69 & 2013 Mar 18 & H214+65 & 2013 Apr 21 & H231+68 & 2013 Feb 15 & H204+73 & 2013 Apr 21 & H226+75 & 2013 Apr 21 \\
L201+75 & 2013 Apr 8  & H220+65 & 2013 Feb 15 & H237+68 & 2013 Feb 15 & H212+73 & 2013 Apr 21 & H235+75 & 2013 Apr 21 \\
L222+75 & 2013 Mar 18 & H226+65 & 2013 Feb 15 & H208+70 & 2013 Apr 21 & H220+73 & 2013 Feb 15 & H245+75 & 2013 Feb 22 \\
L244+75 & 2013 Apr 8  & ~       & ~    & ~       & ~    & ~       & ~    & ~       & ~    \\
\hline
\end{tabular}
\end{table*}

\begin{figure}
\includegraphics[width=\hsize]{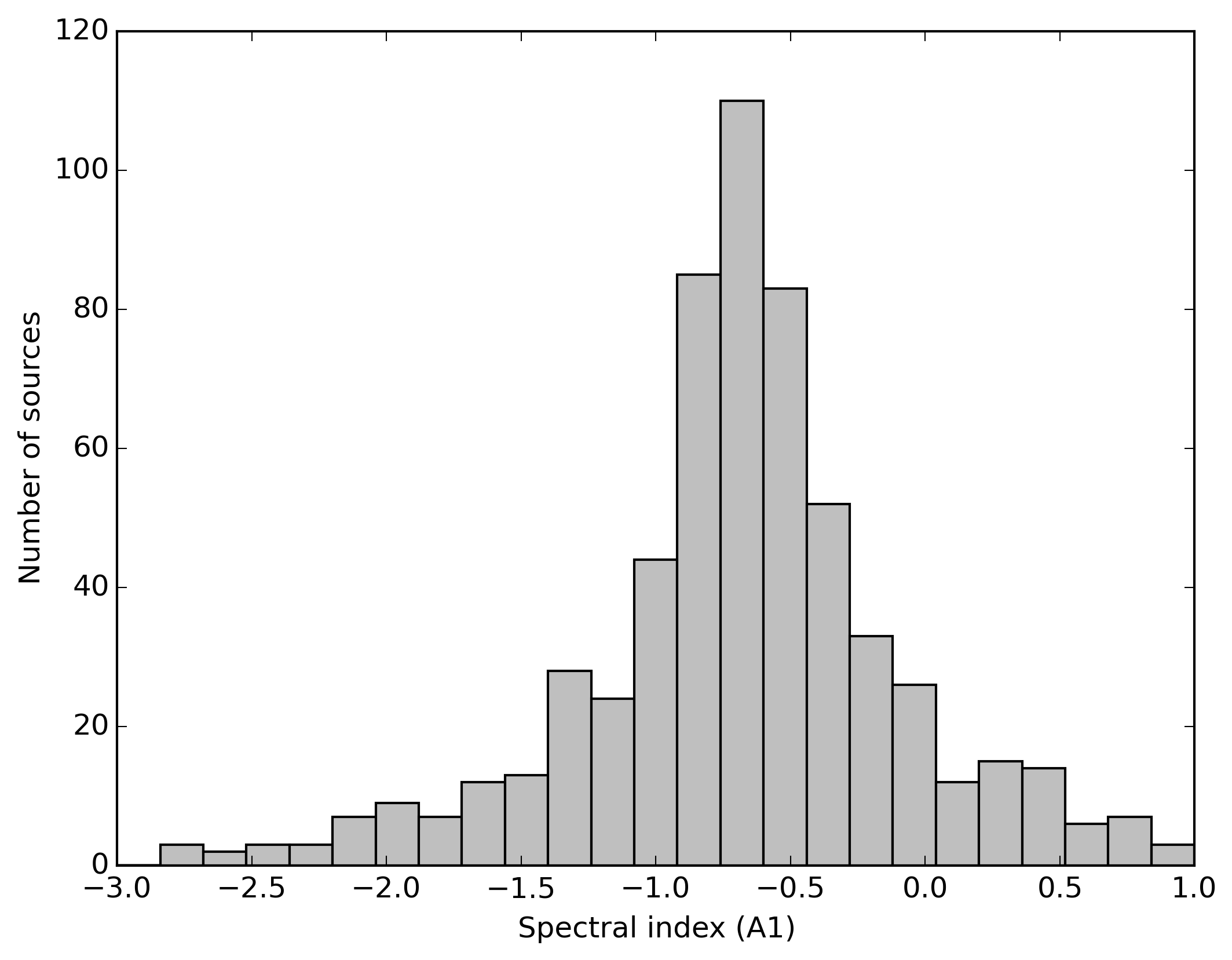}\\
\includegraphics[width=\hsize]{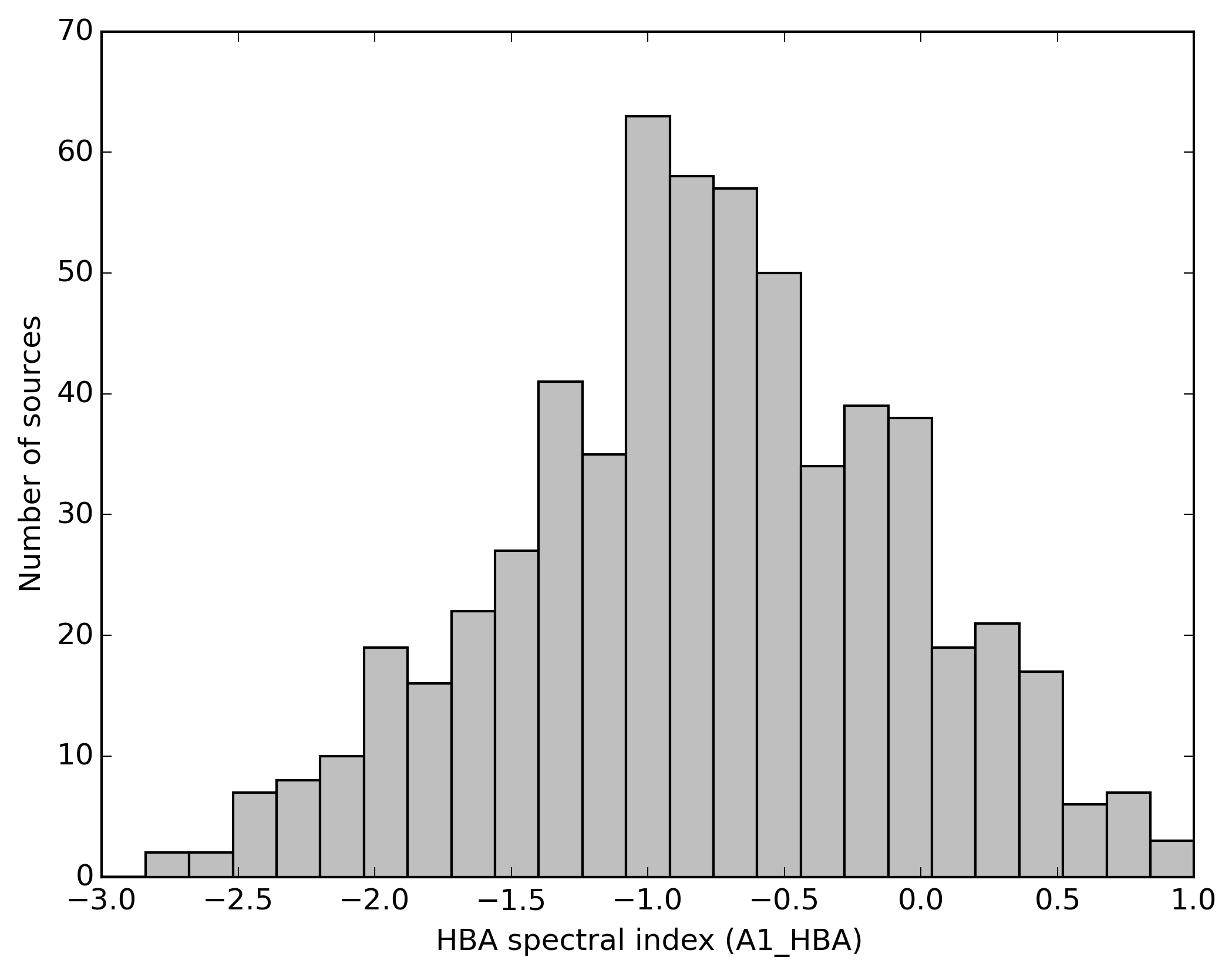}
\caption{Spectral index histograms for the MVF catalog sources with {\tt A0\_HBA}$\,>\,0.2\,\mathrm{Jy}$, based on the cataloged {\tt A1} values ({\it top}), and the same but for HBA values only, catalogued as {\tt A1\_HBA} ({\it bottom}).}
\label{figure:spix}
\end{figure}

We present the spectral index distribution determined on the basis of the MVF catalog in Fig.~\ref{figure:spix}. We consider all sources with a cataloged {\tt A0} value greater than 200 mJy (628 out of 1209 sources). Considering the spectral index determined using all cataloged frequencies (recorded in column {\tt A1}), the mean and median values are $-0.60$ and $-0.66$, respectively. These are somewhat more shallow than spectral indices determined for the same sources from HBA fluxes alone (mean and median of
$-0.70$ and $-0.77$,
respectively). Part of this may be due to spectral turnovers at LBA frequencies or cosmic ray energy loss processes, although some part is likely due to measurement errors. We note that the HBA-only spectral indices have a systematic error due to beam effects not incorporated in the beam model described in \S\,\ref{subsection:imaging}. The error is estimated at the level of
$0.05\pm0.22$
for the MVF region on the basis of more recent electromagnetic simulations of the HBA stations including the effects of mutual coupling. Because these simulations are still under development, and the correction at the high elevation of the MVF snapshots is expected to be very small, we have not adjusted the HBA spectral indices presented in the bottom panel of Fig.~\ref{figure:spix}.
We will fully address this issue during the creation of the full MSSS catalog.

\subsection{Completeness and false detection rate}
\label{sec:completeness}

We determined the completeness of the MVF portion of the survey in the
standard way through the use of
Monte Carlo simulations, injecting simulated point sources into residual
images from the survey and attempting to recover them with the source
finder.
Note that this approach only considers systematic issues related to source identification and characterization in the image plane.
We used PyBDSM for this process; results with PySE would be
expected to be very similar. The only complication arises from the
fact that averaged images (from all LBA and HBA bands) are used as
detection images in the cataloging process. Therefore, 
as a major step in
finding the completeness in individual bands, we replicated the full
procedure used to generate and apply detection images during the source
finding process.

In detail, we generated residual maps from the actual images used for
the cataloging, after first removing any detected sources with
PyBDSM: this ensures that the noise and its spatial distribution are
consistent with that of the real data. We drew simulated sources from a
power-law flux density distribution with $\frac{{\rm d}N}{{\rm d}S}
\propto S^{-1.6}$, with fixed upper and lower flux densities chosen to
span the range of observed flux densities in the survey. For the
multi-band analysis we additionally drew source spectral indices from
a Gaussian distribution with mean $-0.7$ and standard deviation 0.35, and
considered the flux density to be at a reference frequency in the middle
of the band (135 MHz for HBA, 50 MHz for LBA). A suitable number of
simulated sources were then added at random positions to the residual
maps for each band. For the multi-band analysis, we constructed a
detection image by averaging the individual bands (in the case of the
HBA data, this was done by using residual images without beam
correction, taking account of the beam correction factor by scaling
the input fluxes) in order to mimic as closely as possible the process
used in cataloging. Finally, PyBDSM was used to attempt to recover
the simulated sources from the resulting image, taking care to use
exactly the same parameters as applied in the cataloging: a source
was deemed to have been recovered if PyBDSM detected a source within 1
arcmin of the input position with a flux density that matched to
within $10\sigma$, where $\sigma$ is the flux density error returned
by PyBDSM. (In practice, positions normally matched to within less than
one pixel.) This process gave a list of matched and unmatched sources,
which, after repeating several times to improve the statistics, could
be used to estimate the survey completeness.

\begin{figure*}
\centering
\includegraphics[width=0.49\textwidth]{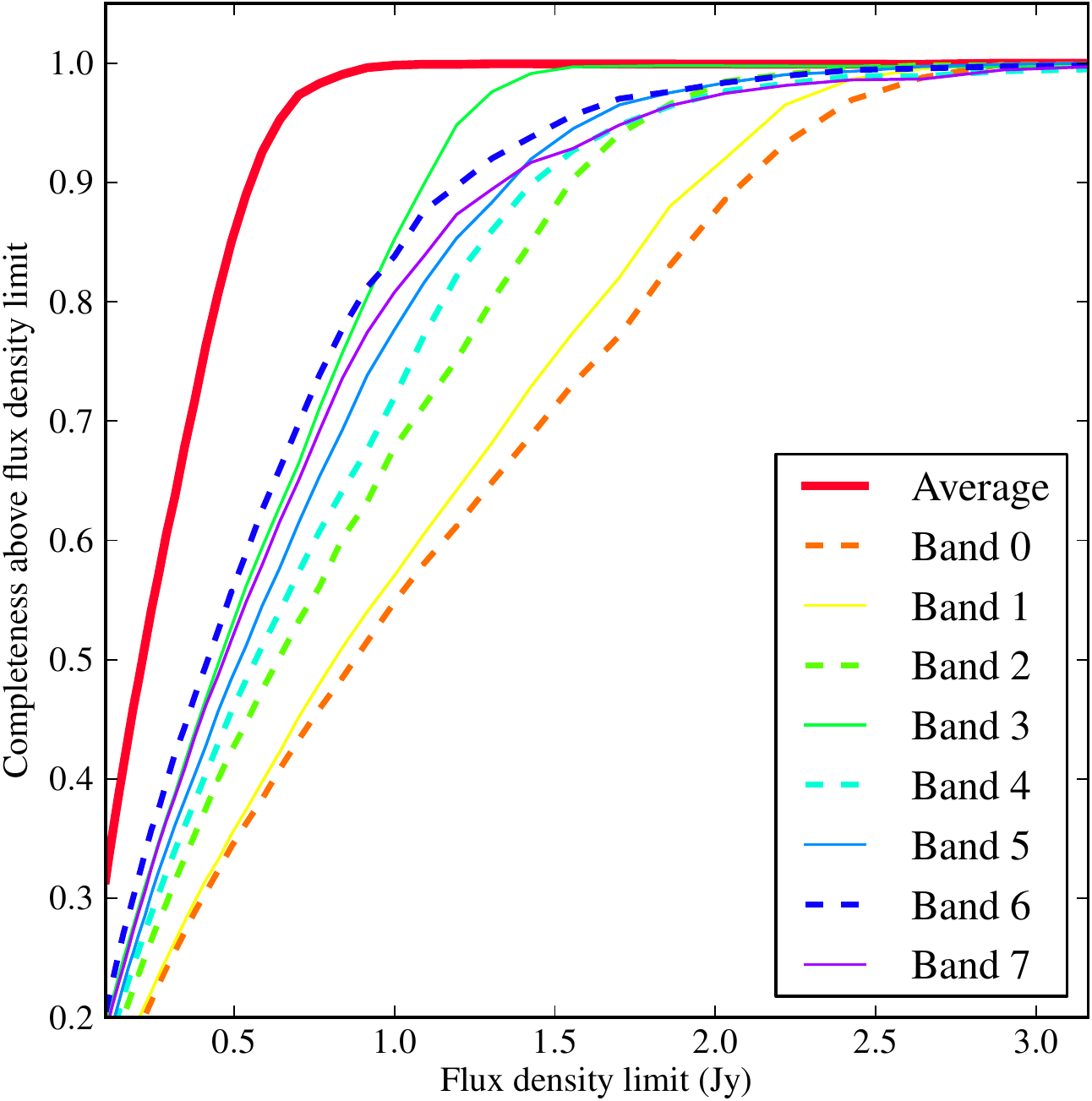}\hfill\includegraphics[width=0.49\textwidth]{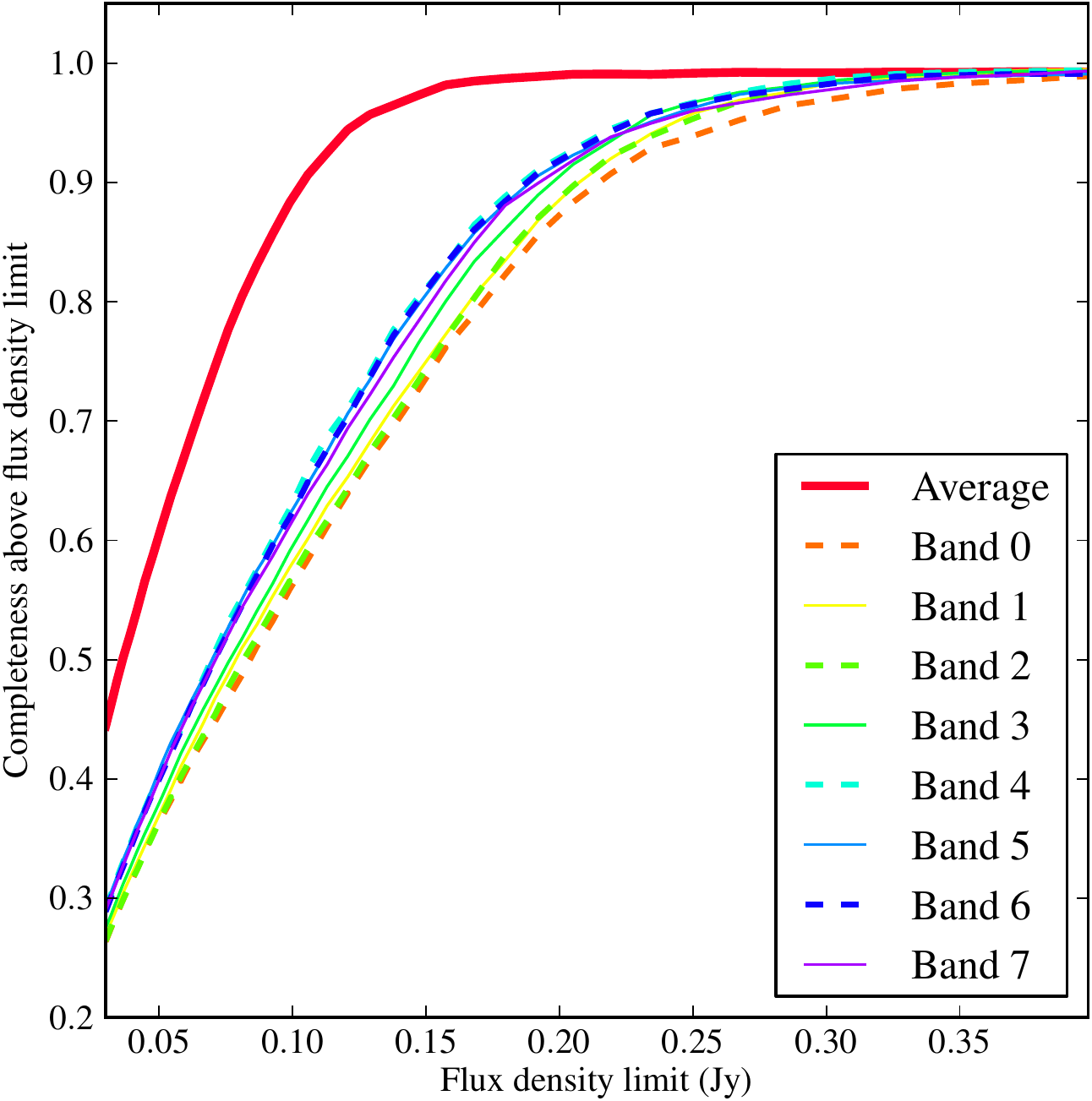}
\caption{Results of completeness simulations for LBA (left) and HBA
  (right). Curves show the cumulative completeness for the individual
  frequencies in the LBA and HBA and for their combination at a
  nominal reference frequency of 50 and 135 MHz, respectively.}
\label{figure:completeness}
\end{figure*}

The results of the completeness simulations for LBA and HBA are shown in Fig.~\ref{figure:completeness}. This figure shows
cumulative completeness curves, i.e. it gives the completeness for
sources above the indicated flux density limit. We see that the HBA catalog is
expected to be 90\% complete above 100 mJy, and 99\% complete by
around 200 mJy at a mid-band reference frequency of 135 MHz; the cataloging process using the averaged
detection image gives, as expected, a catalog that is roughly a
factor $\sqrt{8}$ more sensitive than those derived for the individual
bands. The improvement in sensitivity realized after frequency averaging suggests that the MSSS-HBA images have not yet reached the classical confusion limit (cf. \S\,\ref{subsection:lbasetup}). The LBA catalog has a much higher completeness threshold of
0.55 Jy (90\% complete) or 0.80 Jy (99\% complete) at a reference
frequency of 50 MHz.

We emphasise that these completeness curves are for point sources only
(though at the resolution used that includes nearly all real sources),
that the process assumes that the beam is well modelled as a
Gaussian, and that residual images are free from real structure, which
are good assumptions for the HBA images but much less so for the LBA.
We also assume that there are no relative flux scale offsets within
the LBA and HBA bands. Any departures from these assumptions will tend
to make the real catalogs less complete than indicated by the
completeness curves. At present we regard the HBA curves as reliable,
but the LBA curves should be taken as indicative only.
We note that the simulation used for our completeness estimates
only considers effects present in the image plane. During the development
of the {\tt awimager}, simulations of ideal point sources were used to
demonstrate excellent image plane recovery of objects added to the
visibilities \citep{tasse_etal_2013}, so we do not expect a substantial
effect on the completeness
due to issues in the imaging software that we use. Imperfections
in our calibration solutions and beam model may negatively impact
completeness, but a more detailed simulation to address those effects
is beyond the scope of this paper. A forthcoming paper will present the
MSSS catalog over the full survey area, and with that much larger
statistical sample some of these issues may be more effectively addressed.

\begin{figure*}
\centering
\includegraphics[width=\textwidth]{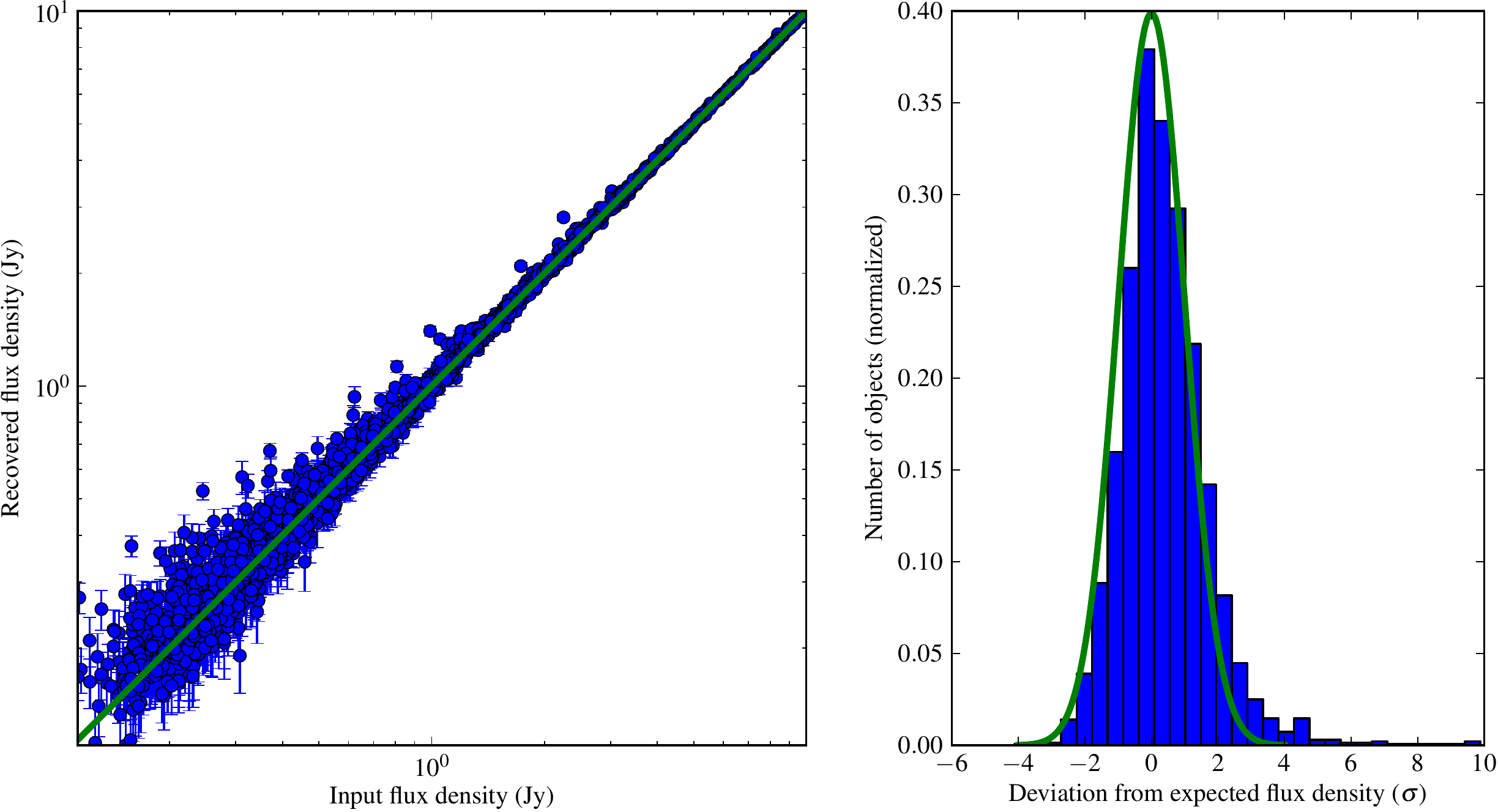}
\caption{Comparison of input and output model source flux densities
  for one band of the HBA (left) and histogram of the deviation from
  the expected value (right). The
  green line on the left shows the line of equality, on the right the
  expected normal distribution if the fluxes are simply affected by
  the noise estimated from the fit.}
\label{figure:completeness-flux}
\end{figure*}

Finally, a by-product of this simulation process is a test of the
reliability of source flux densities as extracted with PyBDSM. We show
a representative plot in Fig.\ \ref{figure:completeness-flux}. It can be
seen that PyBDSM recovers the flux density very accurately. A few
sources at low flux densities are found to have significantly (a
factor of a few) high
flux densities relative to the input values: we attribute this to
confusion (i.e. there is some overlap with a nearby bright source
which is not completely deconvolved by PyBDSM). As the right-hand
panel of Fig.\ \ref{figure:completeness-flux} shows, however, such
sources are a very small fraction of the total.

Having addressed the completeness of the MVF catalog, we now proceed to assess the possibility of falsely detected sources being included in the catalog. To that end we have cross-matched the MVF catalog with existing radio surveys, primarily the deeper 1400\,MHz NVSS catalog. Of the 1209 sources in the MVF catalog, we find that all but 8 are associated with an NVSS source. That result is based on searching for counterparts within the MSSS resolution element and a visual comparison to recover associations with components of extended sources. The 8 remaining MSSS sources were then compared with VLSSr and the 151-MHz 7C survey \citep[][see \S\,\ref{sec:compare7c}]{hales_etal_2007}. Three correspond to VLSSr sources, but none were found in 7C. Thus there are 5 MVF sources that have no associated previously cataloged radio source. All have flux densities between $50-200\,\mathrm{mJy}$ and could be steep spectrum sources below the detection thresholds of both NVSS and VLSSr (a spectral index of $\alpha\approx-2$ would be sufficient), but we nevertheless consider them to represent an upper limit to the MSSS false detection rate (FDR), $\lesssim0.4\%$.

\subsection{{\tt CLEAN} bias}
\label{subsection:cleanbias}

Here we begin to characterize the {\tt CLEAN} bias \citep[see, e.g.,][]{becker_etal_1995} present in the MSSS catalog. As noted in \S\,\ref{subsection:imaging}, we anticipate a small bias because of the excellent $uv$ coverage and our careful deconvolution procedure. To characterize the effect in MSSS data, we have performed a simulation adding artificial sources into the visibilities, followed by an imaging sequence carried out in the same way as for the actual data. We started with the visibility data from all 8 HBA bands of a representative field in the MVF region (H229+70). Real sources that are characterized in the MVF catalog were subtracted in the visibility plane. Artificial point sources were then drawn from a population similar to the one used for the completeness study in \S\,\ref{sec:completeness} ($\frac{{\rm d}N}{{\rm d}S}\propto S^{-1.6}$, here spanning flux densities between $30\,\mathrm{mJy}$ and $30\,\mathrm{Jy}$). These were added into the visibilities, incorporating the LOFAR beam model. Next, the visibility data were imaged as described in \S\,\ref{subsection:imaging}. In particular, an initial shallow clean of each band was used to generate a stacked full-bandwidth image that defines a mask for a subsequent deep deconvolution of each separate band. We used the {\tt awimager} for the imaging steps including the full LOFAR beam model. We also verified our results by injecting the same sources into the visibilities without incorporating the LOFAR beam model and imaging the resulting visibilities in $\textsc{casa}$ with the same settings as were used in {\tt awimager} (e.g. the same {\tt CLEAN} mask).

Pixel values were drawn from the known locations of injected sources and compared with the input flux density values. We followed this procedure to eliminate complications due to completeness effects, which is characterized separately in \S\,\ref{sec:completeness} but would have affected our {\tt CLEAN} bias estimate had we relied on source finding to determine the reconstructed flux of each artificial source. We find a typical {\tt CLEAN} bias of $10\,\mathrm{mJy}$ in each band, well below the typical noise level in each band ($\approx30\,\mathrm{mJy\,beam}^{-1}$) and far below the completeness threshold. Multiple realizations of the same artificial source population were utilized to assess the robustness of the estimate. We found that typical values cluster around $10\,\mathrm{mJy}$, but with substantial variation around the $50\%$ level, depending on the particular realization of the artificial source population. For the MVF, we do not make a {\tt CLEAN} bias correction in the catalog since its exact value is uncertain but very small ($\lesssim5\%$) above the completeness limit per band. In the full MSSS catalog, we will characterize this effect in more detail along with the completeness considerations described in \S\,\ref{sec:completeness}. We have not characterized this effect in the LBA portion of the MVF catalog because it is not yet possible within our direction dependent calibration procedure. It is expected to be a small effect for the same reasons as in the HBA catalog.

As expected, the {\tt CLEAN} bias measured in MSSS-HBA is far lower than in other recent radio surveys. For example, in the original VLSS survey catalog, the bias was characterized as $1.39\sigma$ (variable with the local rms noise level), and substantially improved in VLSSr to $0.66\sigma$. In the same terms, the {\tt CLEAN} bias is $0.67\sigma$ in NVSS and $1.67\sigma$ in FIRST. For MSSS-HBA, we find a {\tt CLEAN} bias of approximately $0.3\sigma$ using this simulation.

\subsection{Comparison with other catalogs}

\subsubsection{HBA}\label{sec:compare7c}

The HBA MSSS catalog can be compared directly to the 151-MHz 7C
survey \citep{hales_etal_2007}. 7C has a resolution of $70 \times 70\,
\mathrm{cosec}(\delta)$ arcseconds at a frequency of 151 MHz, and covers
1.7 sr of the northern sky, including the area of the MSSS described here.

\begin{figure}
\centering
\includegraphics[width=\columnwidth]{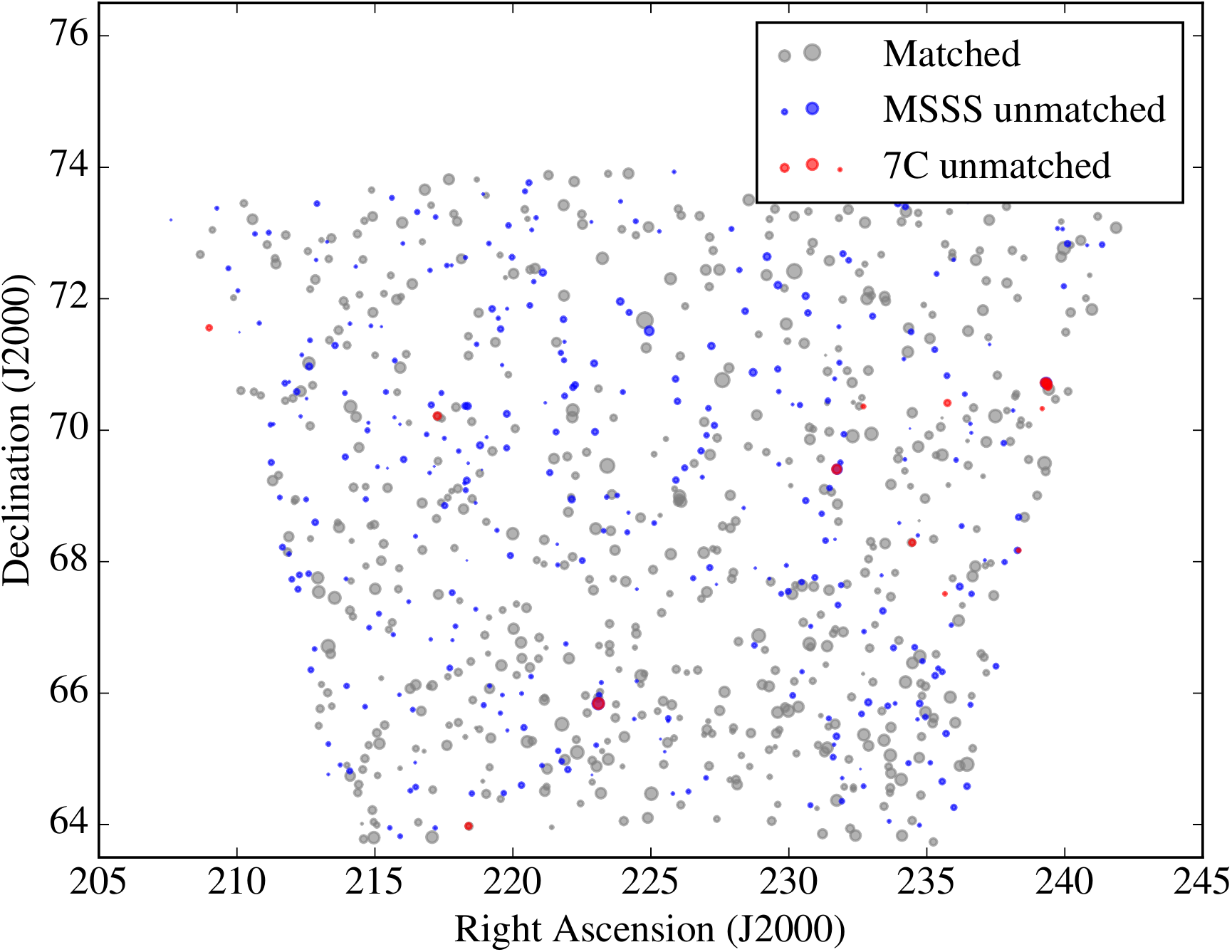}
\caption{Comparison of 7C and MSSS cataloged sources in the MVF.}
\label{figure:7c}
\end{figure}

We filtered the full 7C catalog\footnote{Available electronically
  from \\ {\tt http://www.mrao.cam.ac.uk/facilities/surveys/7c/}} to
produce a catalog within the 100 square degrees of the field described
in this paper, containing 607 7C sources. The completeness limit for
7C in this part of the sky is about 400 mJy, and about 248 7C sources
lie above this completeness limit. We would expect essentially all of
these to be detected by MSSS since, as noted in Section
\ref{sec:completeness}, the MSSS is 99\% complete to 200 mJy at 135
MHz. The rms noise in the 7C images is about 30 mJy, while the noise
in the single-band HBA images is generally less than 20 mJy, and so in
general we expect MSSS to go deeper than 7C. Indeed, after
filtering out cataloged sources with poorly constrained positions and
those without measured 151\,MHz flux densities, there are 1101 MSSS
sources in the area of interest, a factor of 1.8 more than in 7C.

We cross-matched the 7C and MSSS catalogs by combining the
random (not systematic) positional errors of each pair of
sources in quadrature, and finding the maximum-likelihood 7C match for
each MSSS object. Bearing in mind that the distribution of offsets is
a Rayleigh distribution, we imposed a threshold on the acceptable
maximum likelihood, determined by inspection of the matching results,
to reject spurious matches.
This method makes optimal use of the known
positional errors. The initial cross-match resulted in a mean
  positional offset of $7.6 \pm 0.8$ and $-0.8 \pm 0.3$ arcseconds in
  RA and Dec respectively: we removed this offset (which is consistent
  with the offset already determined above) and repeated the
  cross-matching process. A total of 592 sources, or 98\% of the 7C
  sources, matched between the two catalogs, and all but 4 of the 7C
  sources above the completeness limit (98\%) are matched with MSSS sources.

It is useful to look at the positions and flux densities of matched and
unmatched sources (Fig.\ \ref{figure:7c}). We see that most unmatched
sources are faint MSSS sources, which is just a consequence of the
fact that MSSS is more sensitive than 7C over most of the sky area. The
very few bright unmatched 7C sources are almost all positionally nearby
to unmatched MSSS sources of similar brightness, which means that
these are almost certainly the same sources with positions that differ
by too much for the algorithm to consider them to match. Sources of unmodelled
positional uncertainty include the slightly higher resolution of
the 7C survey.

\begin{figure*}
\centering
\includegraphics[width=0.49\linewidth]{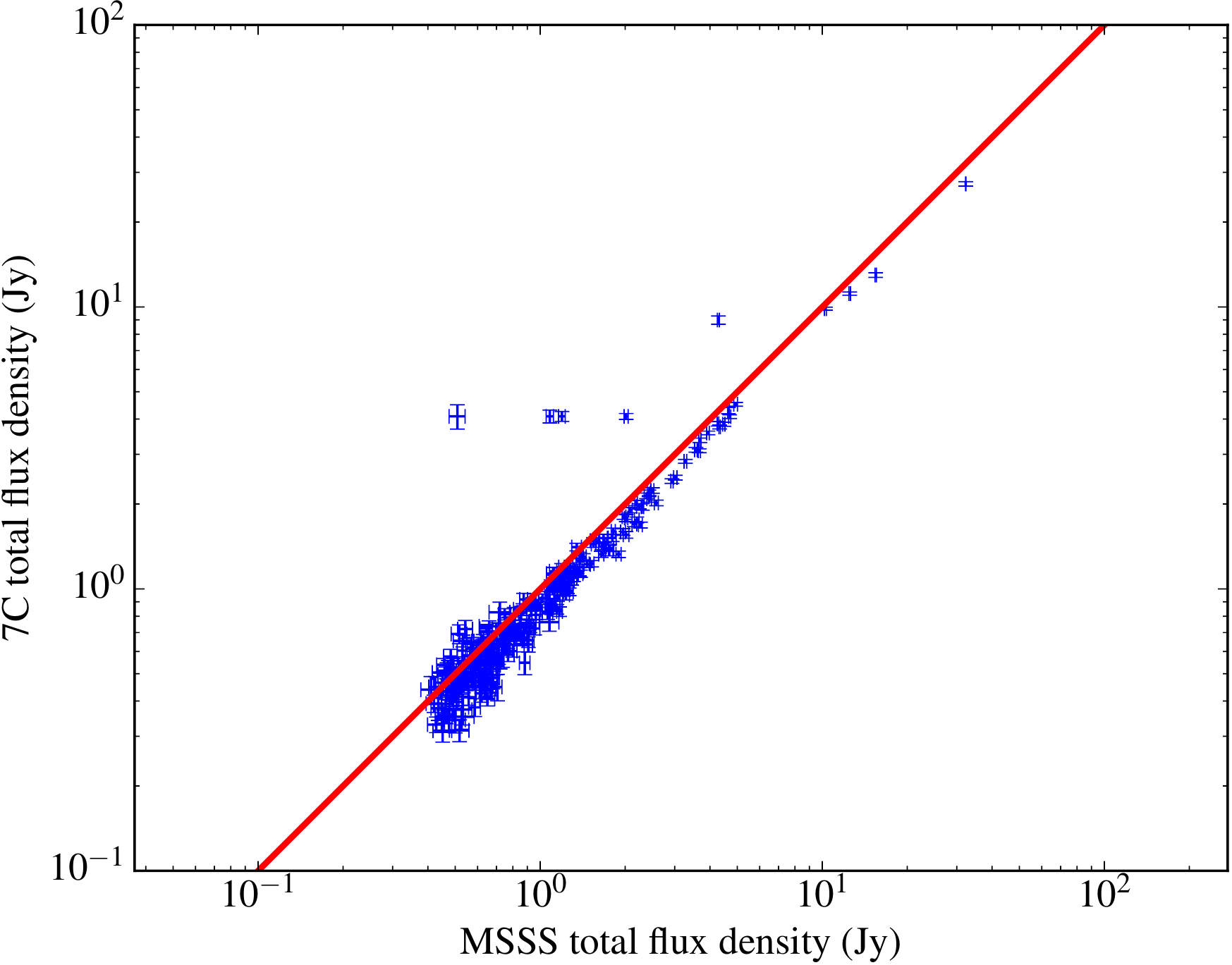}\hfill
\includegraphics[width=0.49\linewidth]{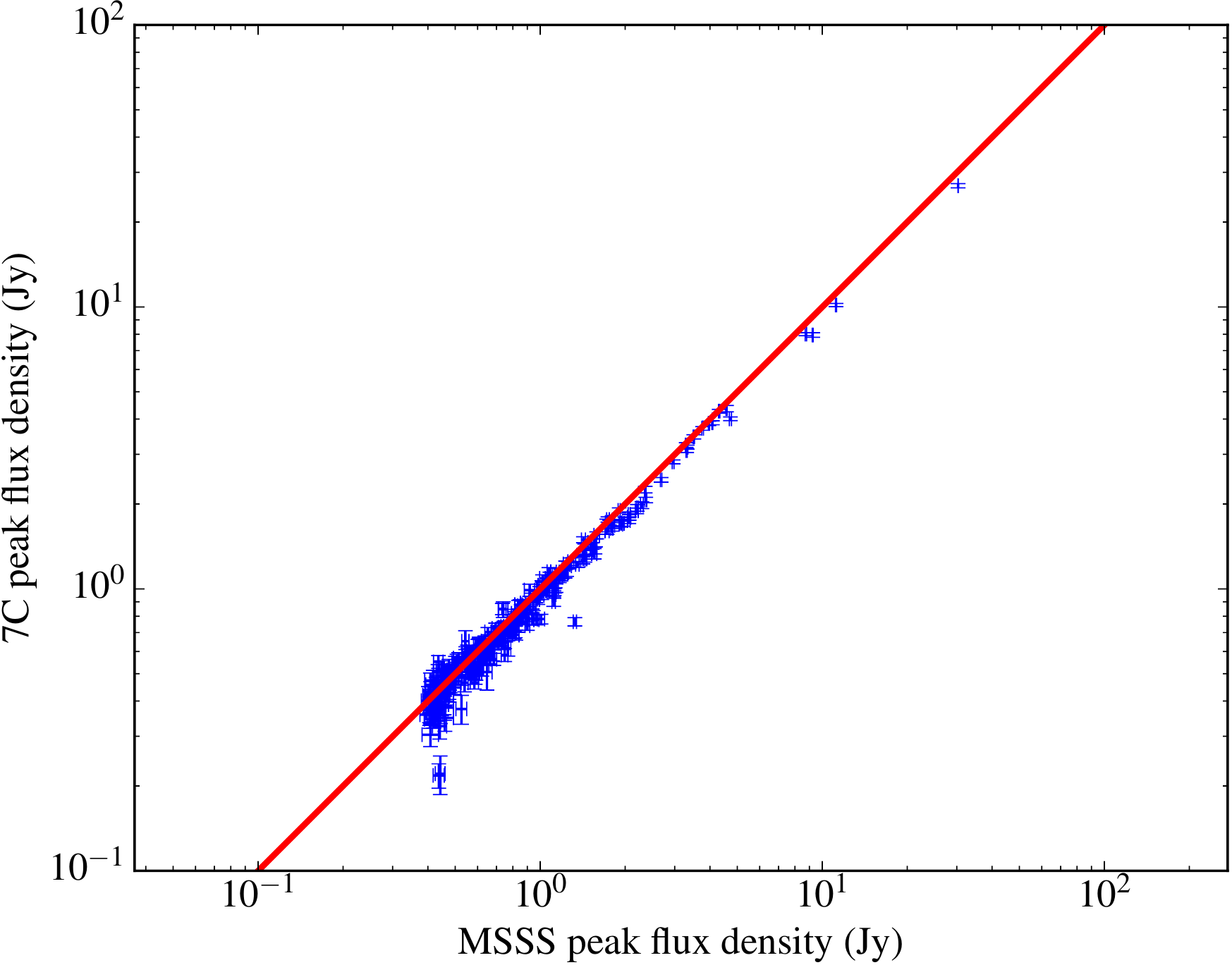}
\caption{Comparison of total ({\it left}) and peak ({\it right}) flux densities for bright
  sources in the MSSS and 7C surveys.}
\label{figure:7c-ff}
\end{figure*}

Finally, we can consider the flux scales of the two catalogs. Here we
consider only the subsample of the catalog above the 7C completeness
limit of 400\,mJy, since the flux densities of parts of the catalog
where the samples are not complete will tend to be biased high. There
is an excellent correlation (Fig.~\ref{figure:7c-ff}) between the flux
densities from the two samples. A very few sources have much larger
total flux densities in 7C compared to MSSS, but these differences are
not present in the peak flux-density analysis, and so must arise from
differences in the algorithms for fitting to extended sources in the
two surveys. We see that there is a slight but clearly significant
difference in both the peak and integrated flux densities in the 7C
and MSSS catalogs, in the sense that the MSSS flux densities are
systematically high by about 9\% (total flux density) or 6\% (peak
flux density). Both surveys should be tied to the flux scale of
\cite{roger_etal_1973}, so in principle we would not expect this
systematic difference. In practice, the difference is likely due to
different assumptions about the flux density of 3C\,295, the reference
flux calibrator for almost all the MSSS observations, which seems
likely to have been the primary flux calibrator for the 7C
observations as well given the RA of the field
\citep{mcgilchrist_etal_1990}. The 7C catalog uses the 6C flux density
of this object, 89.8 Jy, whereas the 150\,MHz flux density
interpolated from the fits of \cite{scaife_heald_2012} is 97.7 Jy,
giving the correct direction and approximately the observed magnitude
of the flux density discrepancy. Indeed, the 150\,MHz flux density of
3C\,295 appears anomalously low on the flux plot of Scaife \& Heald,
compared to, e.g., the 178\,MHz 3CRR flux density. We are therefore
confident in the flux scale and flux recovery in the catalog: further
investigation of the true flux density of 3C\,295 at HBA frequencies
would be desirable.

\subsubsection{LBA}

LBA comparisons are restricted by the relatively small number of
cataloged sources with low-frequency flux densities. However, we have
compared the MSSS MVF with the 8C at 38 MHz and the VLSS at 74 MHz in
the same way as in the previous section. We find that there are a
similar number of MSSS and 8C sources in the field (279 MSSS sources,
233 8C sources), but only around 80\% of MSSS sources (above $\sim
0.8$ Jy at 50 MHz, the 99 per cent completeness level) have a
counterpart in 8C, and similarly there are 8C sources with no MSSS
counterpart. The 8C flux densities are higher than the MSSS ones by
about 11\%, with significant scatter. By contrast, the VLSS appears to
go slightly deeper than MSSS, but at the 99\% completeness level
nearly all (94\%) of the MSSS sources have VLSS counterparts, with a
good agreement between peak fluxes -- the VLSS sources are brighter by
about 3\%, with little scatter. A detailed comparison with these
catalogs is deferred until a larger field is available.
We note that while the MSSS data will ultimately allow a deeper catalog, the current images are limited by substantial ionospheric errors that are only partially corrected  with our direction dependent calibration and imaging procedure. Moreover the LBA field is composed of just a single pointing, limiting the full sensitivity to the center of the $10\degr\times10\degr$ region considered here. We expect the final LBA data products to provide substantially deeper levels than both the 8C and VLSSr surveys.

\section{Scientific capability}\label{section:science}

To assess how MSSS compares with other existing, ongoing, and future surveys, we list key parameters in Table \ref{table:surveys}. The survey parameters illustrate that MSSS is complementary to existing catalogs, with potential for even better image quality with followup reprocessing steps.

\begin{table*}
\caption{Nominal MSSS parameters and comparison with other surveys}
\label{table:surveys}
\centering
\begin{tabular}{lllll}
\hline\hline
Survey & Frequency & Sensitivity & Resolution & Area \\
\hline
MSSS-LBA & 30--78~MHz   & $\lesssim50$~mJy~beam$^{-1}$ & $\lesssim150\arcsec$ & 20\,000 $\Box\degr$ ($\delta>0\degr$) \\
8C       & 38~MHz        & $200-300$~mJy~beam$^{-1}$    & $4.5\arcmin\times4.5\arcmin\csc(\delta)$ & \phantom{0}3\,000 $\Box\degr$ ($\delta>+60\degr$) \\
VLSS     & 74~MHz            & 100~mJy~beam$^{-1}$      & $80\arcsec$          & 30\,000 $\Box\degr$ ($\delta>-30\degr$) \\
MSSS-HBA & 120--170~MHz & $\lesssim$\,10--15~mJy~beam$^{-1}$  & $\lesssim120\arcsec$ & 20\,000 $\Box\degr$ ($\delta>0\degr$) \\
7C       & 151~MHz           & 20~mJy~beam$^{-1}$       & $70\arcsec\times70\arcsec\csc(\delta)$ & \phantom{0}5\,500 $\Box\degr$ (irregular coverage) \\ 
TGSS     & 140--156~MHz     & 7--9~mJy~beam$^{-1}$    & $20\arcsec$          & 32\,000 $\Box\degr$ ($\delta>-30\degr$) \\
WENSS    & 330~MHz           & 3.6~mJy~beam$^{-1}$      & $54\arcsec\times54\arcsec\csc(\delta)$ & 10\,000 $\Box\degr$ ($\delta>+30\degr$) \\
NVSS     & 1400~MHz          & 0.45~mJy~beam$^{-1}$     & $45\arcsec$          & 35\,000 $\Box\degr$ ($\delta>-40\degr$) \\
\hline
\end{tabular}
{\\{\it Note. Sensitivity and resolution values for the MSSS survey components are upper limits corresponding to images produced with baselines shorter than $3\,\mathrm{k}\lambda$. Longer baselines are included in the observations as a matter of course, enabling reprocessing toward the production of an updated, higher angular resolution catalog.}}
\end{table*}

\begin{figure*}
  \includegraphics[width=0.49\hsize]{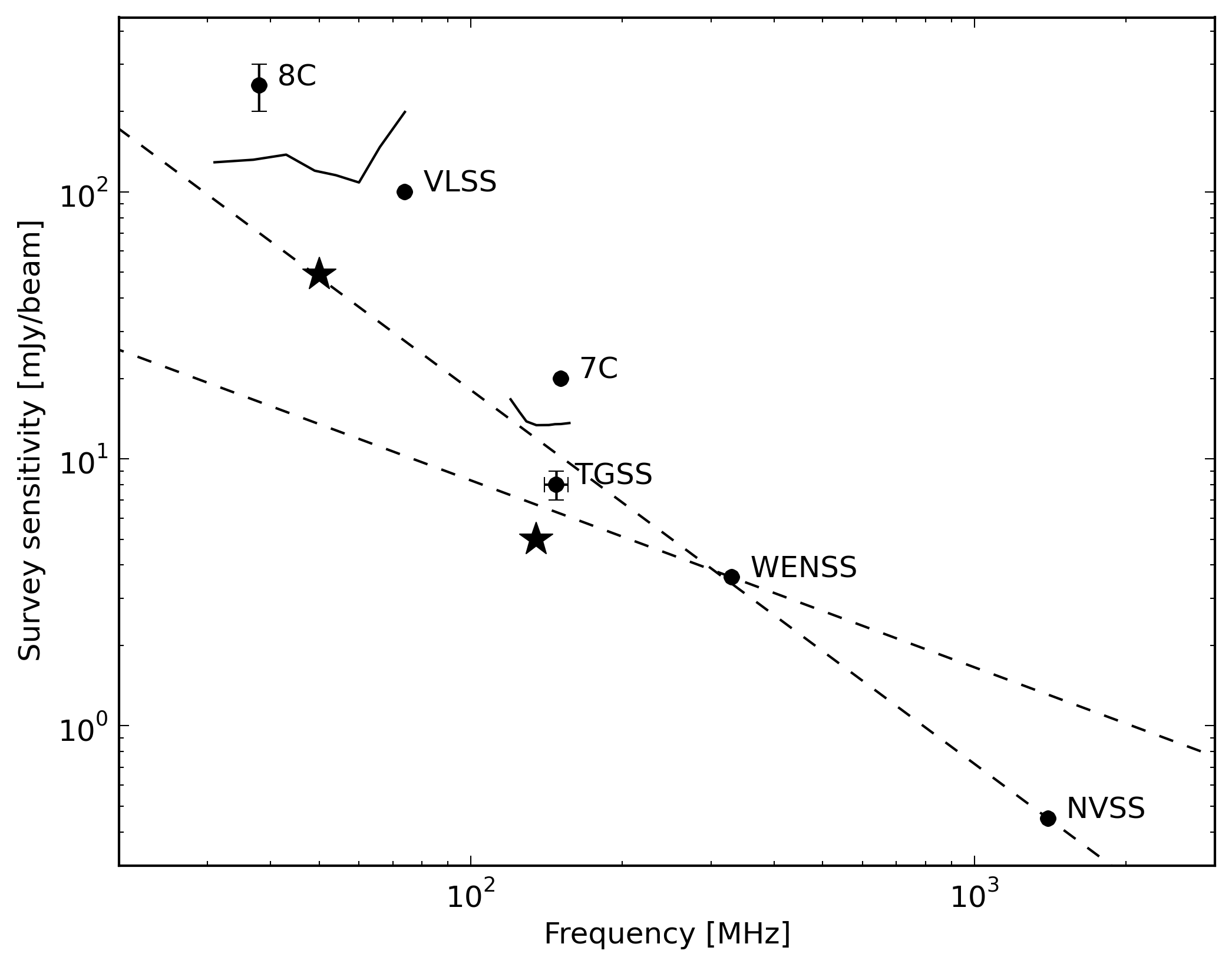}\hfill\includegraphics[width=0.49\hsize]{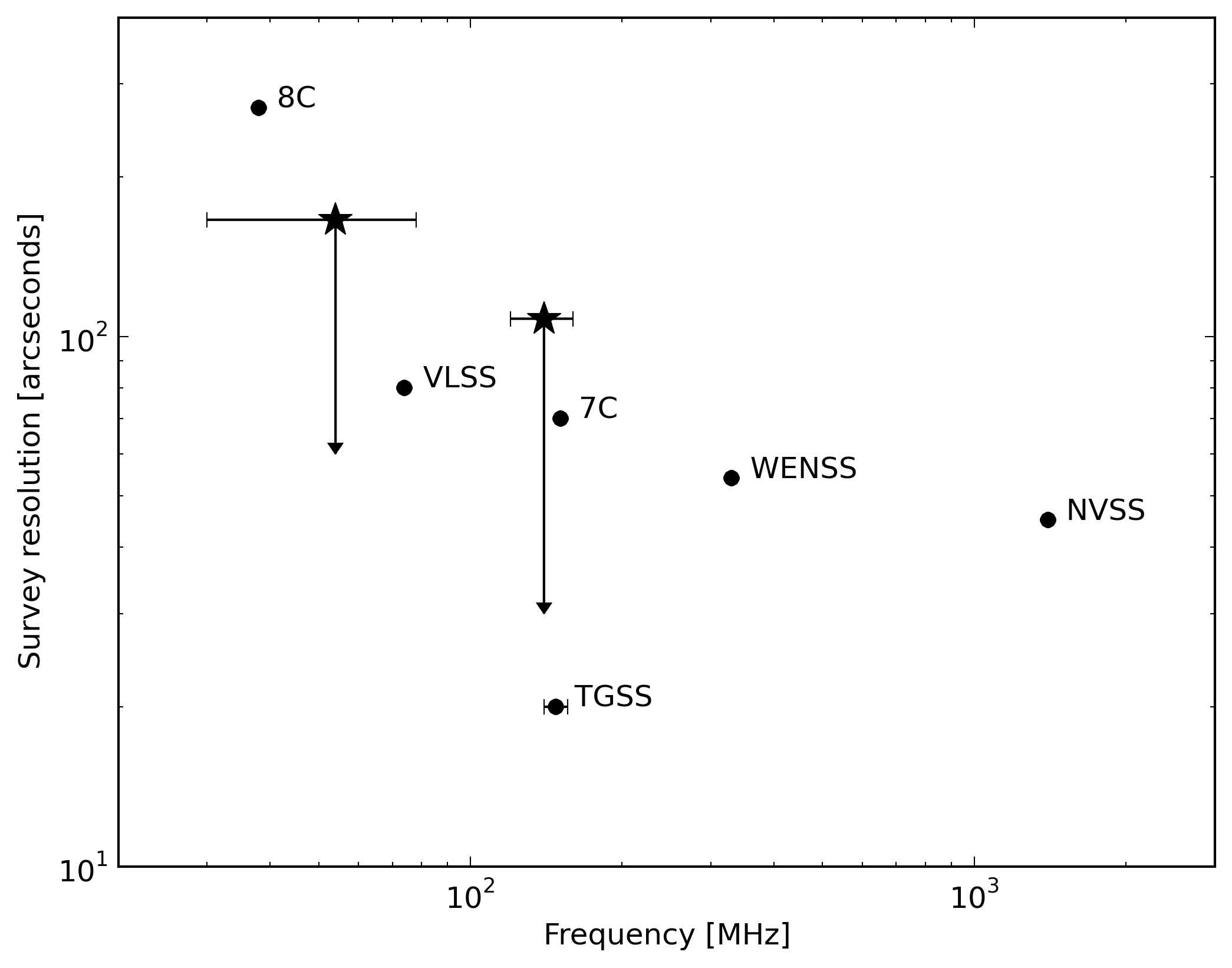}
  \caption{Comparisons between MSSS sensitivities ({\it left}) and resolutions ({\it right}) with those of other existing radio surveys as summarized in Table~\ref{table:surveys}. In the lefthand panel, dashed lines indicate representative spectral indices of $\alpha=-0.7$ and $\alpha=-1.4$. The solid black lines illustrate the frequency dependence of the sensitivity in the 8 bands in each of the LBA and HBA survey segments, while the black stars show the frequency-averaged sensitivity demonstrated in \S\,\ref{section:mvf}. In the righthand panel, the downward-pointing arrows indicate that the angular resolution of the initial MSSS catalog is limited with respect to the capabilities of the visibility data. Processing the full array will improve the survey performance.}
  \label{figure:msss-compare}
\end{figure*}

The values for sensitivity and resolution are shown graphically in Figure \ref{figure:msss-compare}. The sensitivity panel shows that sources with typical spectral indices will be detected in both WENSS and MSSS. Steep spectrum sources detected in NVSS will also be recovered in MSSS. Therefore, the sensitivity of MSSS make it interesting in terms of comparison with other all-sky surveys. The study of steep spectrum sources (e.g., radio galaxies and diffuse synchrotron sources) in particular will be strongly enhanced by MSSS. It is clear from Fig.~\ref{figure:msss-compare}, however, that a larger LOFAR array should be processed in order for the MSSS angular resolution to be competitive with the other existing surveys.

\subsection{Long baselines}\label{subsection:hires}

A uniquely powerful aspect of LOFAR's view of the low frequency sky is the sub-arcsecond resolution afforded by baselines to and between international stations \citep[see, e.g.,][]{varenius_etal_2015}. International stations are always included in the LBA observations, but not in the HBA observations for the reasons described in \S\,\ref{subsection:hbasetup}. A long-baseline working group is planning to use the same MSSS data that will generate the initial low-resolution images and catalog as part of a survey for long-baseline calibrators \citep[see also][]{moldon_etal_2015}, and to provide initial images for the brightest sources. Test processing rounds have demonstrated that higher angular resolution imaging is feasible despite the sparse $uv$ coverage. HBA images at 5--30\arcsec resolution (using the outer Dutch remote stations) have been successfully produced. Efforts toward a large-scale reprocessing of the MSSS data to produce a high-resolution catalog are in an early phase at the time of writing.

\begin{figure*}
\centering
\includegraphics[width=0.8\hsize]{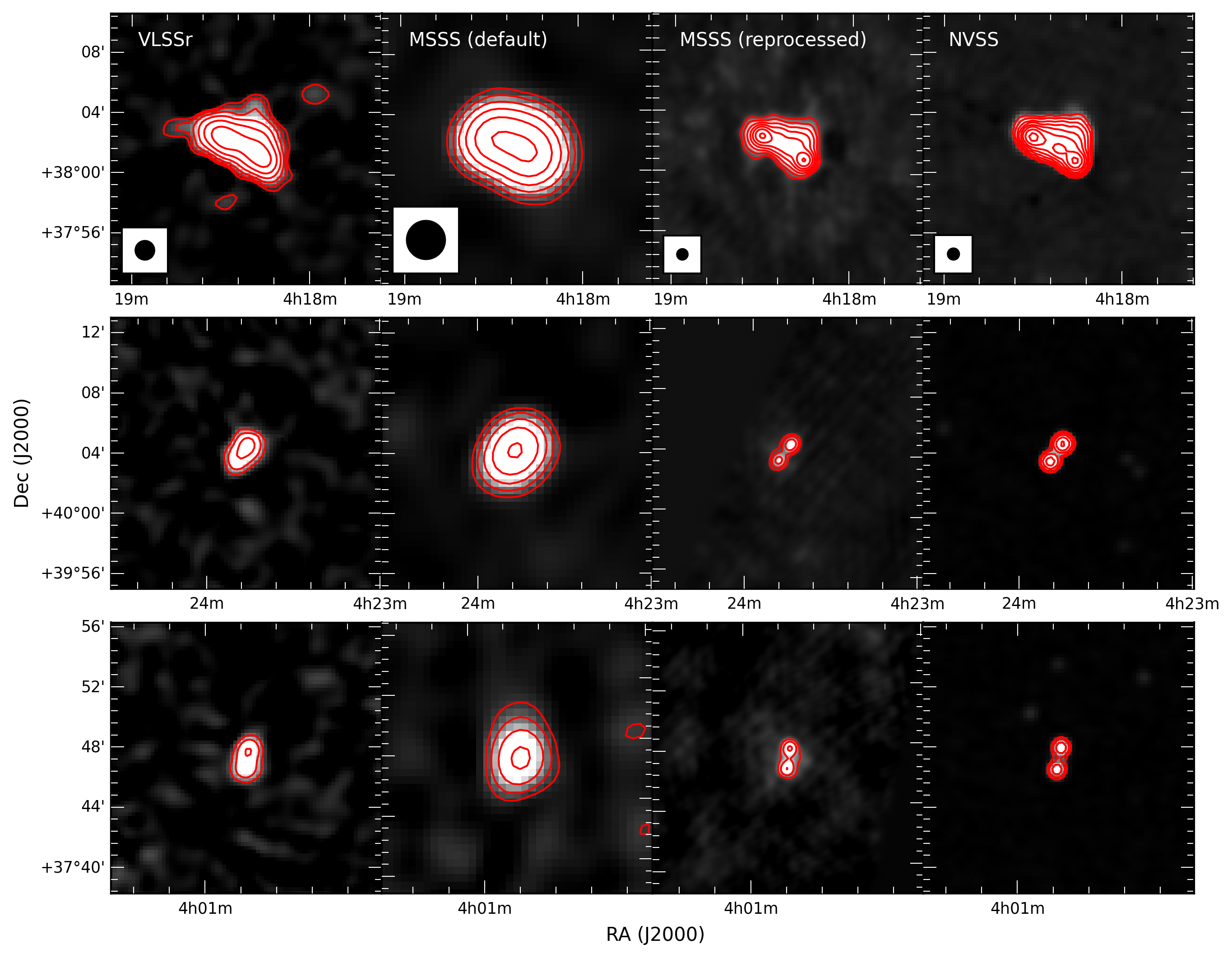}
\caption{Images of three sources, one per row, identified in the field of H063+39. The sources are 3C~111 ({\it top}), 4C~+39.13 ({\it middle}), and 4C~+37.10 ({\it bottom}). In each row the left panel displays the VLSSr image (resolution $75\arcsec$), the next two columns display MSSS images at $2.5\arcmin$ and $42\arcsec$ resolution (see \S\,\ref{subsection:hires}), and the right panel displays the NVSS image (resolution $45\arcsec$). The beamsize for each column is plotted in the top row. Contour levels begin at $0.6\,\mathrm{Jy\,beam^{-1}}$ for VLSS; 1, 0.2 and $0.2\,\mathrm{Jy\,beam^{-1}}$ for low-resolution MSSS; 0.2, 0.15 and $0.1\,\mathrm{Jy\,beam^{-1}}$ for high-resolution MSSS; and $0.02\,\mathrm{Jy\,beam^{-1}}$ for NVSS. Contours increase by multiples of 2 in all frames.}
\label{figure:hires}
\end{figure*}

An example of higher resolution imaging of MSSS data is presented in Fig.~\ref{figure:hires}. The MSSS-HBA observations of field H063+39 were reprocessed using an automated self-calibration loop that has been developed as part of an overall upgrade to the TIP (see \S\,\ref{subsection:tip}). This scheme performs a progressively higher angular resolution sequence of source finding, phase calibration and imaging. Here, we have performed 8 iterations to achieve a final image resolution of $42\arcsec$ after averaging over 5 bands (cf. the initial resolution of MSSS images, $2.5\arcmin$ from baselines $\leq2\,\mathrm{k}\lambda$; see \S\,\ref{subsection:imaging}). While the data provide the capability for still higher angular resolution, our aim here is to demonstrate image quality comparable to existing higher frequency radio surveys. We have selected 3 sources from the field of view that are resolved in the high resolution MSSS image: 3C~111 and two double-lobed radio galaxies (4C~+39.13 and 4C~+37.10). Through comparison with the NVSS images of the same objects it is clear that the morphology is properly reproduced in the reprocessed MSSS data, at an order of magnitude lower frequency.

\subsection{Transients and variable sources}\label{subsection:transients}

There are multiple aspects of the MSSS survey that are useful for conducting transient searches. First, the 9 (2) snapshots in LBA (HBA) mode allow a limited variability and transient capability. Furthermore, in LBA mode a single subband beam is always placed on the north celestial pole (NCP), enabling transient monitoring on longer timescales (albeit with reduced sensitivity). A full description of the NCP transient monitoring program, and the first low-frequency transient identified therein, is presented in a companion paper \citep{stewart_etal_2015}.

MSSS is also expected to be useful to detect pulsars that are highly scattered to a degree that causes the pulse profile to be indistinguishable in beamformed observations \citep[see][for a description of the beamformed pulsar observing method]{stappers_etal_2011}. In imaging observations, the total flux density from the pulsar will still be visible, permitting study of the low-frequency spectral behavior of such objects.

\subsection{Magnetism}

One of the fundamental scientific themes being pursued using LOFAR is the study of cosmic magnetism \citep{beck_etal_2013}. A magnetism working group is planning to perform Faraday rotation measure (RM) Synthesis \citep{brentjens_debruyn_2005} on MSSS data in order to search for polarized sources, primarily Galactic \citep[pulsars and diffuse foreground emission; see][]{jelic_etal_2014}, but with hopes of detecting polarized active galactic nuclei (AGNs) \citep[e.g.,][]{mulcahy_etal_2014} and giant radio galaxies (GRGs) as well.

\begin{figure}
\includegraphics[width=\hsize]{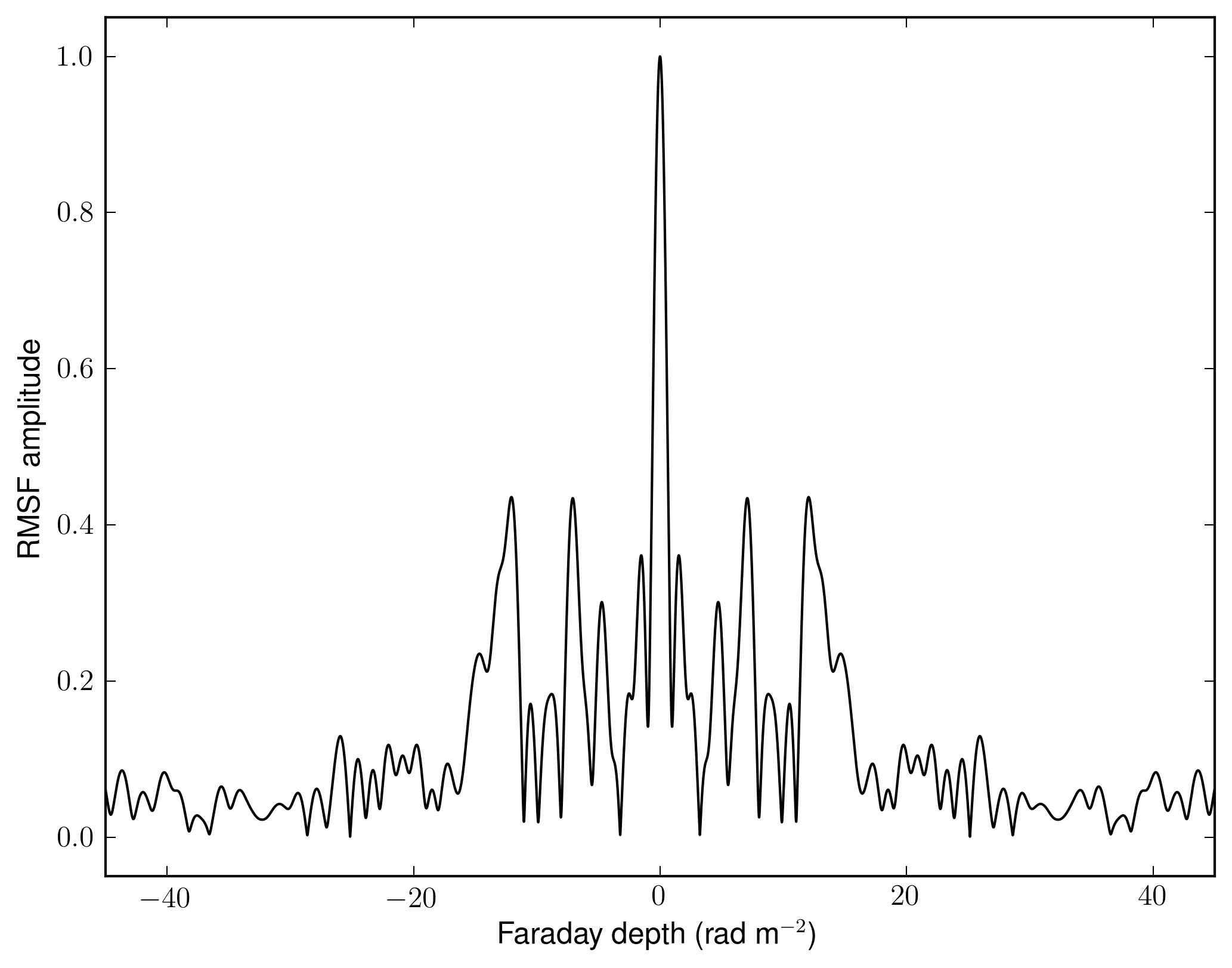}
\caption{Rotation measure spread function provided by MSSS, considering the combination of all 8 HBA bands. The width of the main lobe is $\approx1.3\,\mathrm{rad\,m^{-2}}$, excellent for recovery of small RMs and thus weak magnetic fields.}
\label{figure:rmsf}
\end{figure}

A polarization survey based on MSSS data will be uniquely powerful at these frequencies; with the full achievable angular resolution and sensitivity over the entire survey area, it will help greatly in furthering our understanding of polarization at low frequencies. The HBA component of MSSS provides an excellent Faraday resolution of $\approx1.3\,\mathrm{rad\,m^{-2}}$ and a maximum Faraday depth of approximately $330\,\mathrm{rad\,m^{-2}}$. The rotation measure spread function (RMSF), which displays the response to simple polarized sources using all 8 HBA bands together from MSSS, can be seen in Fig.~\ref{figure:rmsf}. Initial tests of the polarization recoverable through MSSS have already been performed. Polarized Galactic foreground emission from the ``Fan region'' has been detected, matching structures seen in previous observations with WSRT \citep[e.g.,][]{iacobelli_etal_2013}. In addition to this, the highly polarized pulsar PSR~J0218+4232 with a known Faraday depth of $-61\,\mathrm{rad\,m^{-2}}$ \citep{navarro_etal_1995} was detected at the correct Faraday depth after correction for ionospheric Faraday rotation \citep{sotomayor_etal_2013}. This demonstrates that an accurate polarization survey of the Galactic foreground and extragalactic background sources is feasible with MSSS-HBA. This effort will be presented in a future paper.

\subsection{Individual extragalactic and Galactic targets}

The key unique aspect of MSSS is its intrinsic broadband nature that enables study of the spectral properties in various classes of sources. The primary sources of interest include galaxy clusters, star-forming galaxies, AGNs, and supernovae, among others. Several working groups have been formed to investigate preliminary samples of targets and determine their low frequency spectra. Particular scientific questions to be addressed include the spectral properties of galaxy cluster halos and the search for previously unknown diffuse emission; the spectral behavior of nearby star-forming galaxies and the prevalence of spectral turnovers at low frequencies; characterization of GRGs and the search for previously unknown objects; and the search for new supernova remnants and Pulsar Wind Nebulae (PWNe).

\section{Summary and outlook}\label{section:conclusion}

We have presented the motivation and setup of the Multifrequency Snapshot Sky Survey (MSSS), a groundbreaking new radio continuum survey being performed with LOFAR. The primary design goal is a moderate angular resolution ($\sim\,2\arcmin$), moderate depth ($\sim10\,\mathrm{mJy\,beam}^{-1}$), but intrinsically broadband survey to populate the initial LOFAR source database that will be used for calibration purposes in forthcoming deeper observations. The survey is also fertile ground for new scientific research into the source population of the low-frequency sky, and in particular uniquely enables the study of the spectral characteristics of the brighter sources detected by MSSS.

The quality of the forthcoming initial MSSS data release has been illustrated within the MSSS Verification Field (MVF), a representative 100 square degree area centered at $(\alpha,\delta)=(15^\mathrm{h},69\degr)$. We find that the survey area studied here is 90\% complete above 100 mJy in the HBA with angular resolution $108\arcsec$, and above 550 mJy in the LBA with $166\arcsec$ resolution. Source positions and flux densities match well with existing radio surveys in these frequency ranges, albeit with possible small flux scale offsets that require confirmation over a larger sample area.

Subsequent MSSS data releases are anticipated, with the potential for improved multi-directional calibration (essential especially for the LBA portion of MSSS) and higher angular resolution imaging.
These will be accompanied by a number of papers focusing on specific science projects enabled by the MSSS catalog (as outlined in \S\,\ref{section:science}) such as low-frequency transients, polarized sources, the spectral properties of nearby star-forming galaxies, and steep spectrum radio galaxies.

\begin{acknowledgements}
We thank the anonymous referee for comments that have improved this paper. LOFAR, the Low Frequency Array designed and constructed by ASTRON, has facilities in several countries, that are owned by various parties (each with their own funding sources), and that are collectively operated by the International LOFAR Telescope (ILT) foundation under a joint scientific policy.
We acknowledge financial support from the BQR program of Observatoire de la C\^ote d'Azur and from the French A.S. SKA-LOFAR.
C.F. and G.M. acknowledge financial support by the ``{\it Agence Nationale de la Recherche}'' through grant ANR-09-JCJC-0001-01.
D.C., Y.C., A.J.vdH, A.R., and J.S. acknowledge support from the European Research Council via Advanced Investigator Grant no. 247295 (PI: R. Wijers).
R.P.B., J.B., T.H., A.J.S., M.P., and R.F. were funded by a European Research Council Advanced Grant 267697 ``4 Pi Sky: Extreme Astrophysics with Revolutionary Radio Telescopes'' (PI: R. Fender).
D.M., B.A., M.R.B., A.B., F.dG., A.H., M.I., K.S., and C.S. were supported by DFG Research Unit FOR1254 ``Magnetisation of Interstellar and Intergalactic Media: The Prospects of Low-Frequency Radio Observations''.
\end{acknowledgements}

\bibliographystyle{aa}
\bibliography{msss}

\end{document}